\documentclass[sigconf]{acmart}
\usepackage{geometry}
\usepackage{subfigure}
\usepackage{multirow}
\usepackage{cleveref}
\usepackage{xspace}

\usepackage{algorithm} 
\usepackage{algorithmic}

\crefname{figure}{Figure}{Figures}
\crefname{appendixfigure}{Figure A.}{Figures A.}

\crefname{table}{Table}{Tables}
\crefname{appendixtable}{Table A.}{Tables A.}

\usepackage{url}
\usepackage{hyperref}
\usepackage{balance}

\newcommand{\noind}[0]{\noindent}
\newcommand{\noindpar}[1]{\noind {\bf #1}}

\def\etal{\textit{et al.}\xspace}

\newcommand{\changed}[1]{\textcolor{black}{#1}}

\RequirePackage{color}\definecolor{RED}{rgb}{1,0,0}\definecolor{BLUE}{rgb}{0,0,1} 

\AtBeginDocument{%
  }

\copyrightyear{2024} 
\acmYear{2024} 
\setcopyright{rightsretained} 
\acmConference[CHI '24]{Proceedings of the CHI Conference on Human Factors in Computing Systems}{May 11--16, 2024}{Honolulu, HI, USA}
\acmBooktitle{Proceedings of the CHI Conference on Human Factors in Computing Systems (CHI '24), May 11--16, 2024, Honolulu, HI, USA}
\acmDOI{10.1145/3613904.3642555}
\acmISBN{979-8-4007-0330-0/24/05}

\begin{document}

\title[WheelPose: Data Synthesis to Improve Wheelchair User Pose Estimation]{WheelPose: Data Synthesis Techniques to Improve Pose Estimation Performance on Wheelchair Users}

\author{William Huang}
\orcid{0000-0001-7651-2190}
\affiliation{%
  \institution{University of California, Los Angeles}
  \city{California}
  \country{USA}
}
\email{william.huang@ucla.edu}

\author{Sam Ghahremani}
\orcid{0009-0007-5844-4834}
\affiliation{%
  \institution{University of California, Los Angeles}
  \city{California}
  \country{USA}
}
\email{samg2024@berkeley.edu}

\author{Siyou Pei}
\orcid{0000-0003-3802-8298}
\affiliation{%
  \institution{University of California, Los Angeles}
  \city{California}
  \country{USA}
}
\email{sypei@g.ucla.edu}

\author{Yang Zhang}
\orcid{0000-0003-2472-6968}
\affiliation{%
  \institution{University of California, Los Angeles}
  \city{California}
  \country{USA}
}
\email{yangzhang@ucla.edu}

\renewcommand{\shortauthors}{Huang, et al.}

\def\systemname {\textit{WheelPose}\xspace}

\begin{abstract}
 
Existing pose estimation models perform poorly on wheelchair users due to a lack of representation in training data. We present a data synthesis pipeline to address this disparity in data collection and subsequently improve pose estimation performance for wheelchair users. Our configurable pipeline generates synthetic data of wheelchair users using motion capture data and motion generation outputs simulated in the Unity game engine. We validated our pipeline by conducting a human evaluation, investigating perceived realism, diversity, and an AI performance evaluation on a set of synthetic datasets from our pipeline that synthesized different backgrounds, models, and postures. We found our generated datasets were perceived as realistic by human evaluators, had more diversity than existing image datasets, and had improved person detection and pose estimation performance when fine-tuned on existing pose estimation models. Through this work, we hope to create a foothold for future efforts in tackling the inclusiveness of AI in a data-centric and human-centric manner with the data synthesis techniques demonstrated in this work. Finally, for future works to extend upon, we open source all code in this research and provide a fully configurable Unity Environment used to generate our datasets. In the case of any models we are unable to share due to redistribution and licensing policies, we provide detailed instructions on how to source and replace said models. All materials can be found at \url{https://github.com/hilab-open-source/wheelpose}.

\end{abstract}




\begin{teaserfigure}
  \centering
  \includegraphics[width=0.95\linewidth]{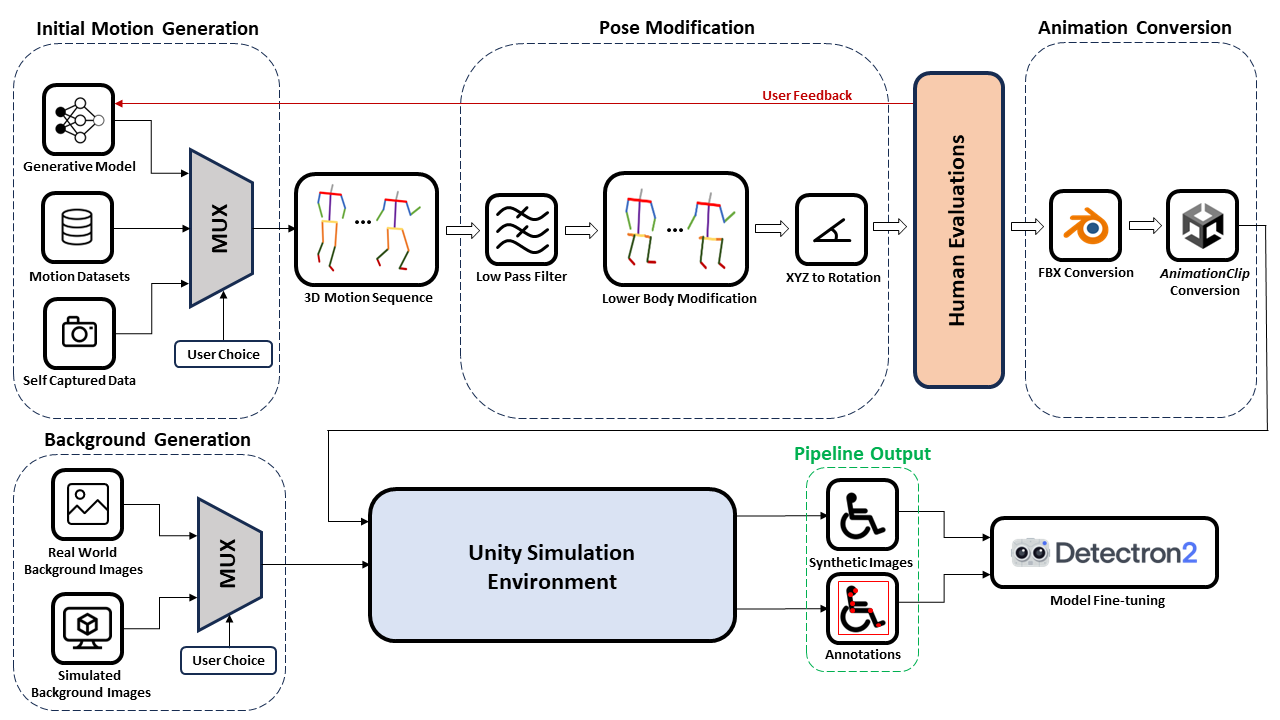}
  \caption{Overview of the full \systemname{} data generation pipeline. Developers can choose different motion sources. Motion sequences are modified according to the specification stated below before being evaluated by human evaluators. Developers can regenerate, filter, and clean motion sequences from human evaluations before all motions are converted into Unity readable \textit{AnimationClips}. Converted motion sequences, selected background images, and parameters are used in simulation to generate synthetic images and their related annotations for use in model boosting.}
  \Description{Eight components: Initial Motion Generation, Pose Modification, Human Evaluations, Animation Conversion, Background Generation, Unity Simulation Environment, Pipeline Output, and Model Boosting. Initial motion generation consists of a choice between generative models, motion datasets, or self captured data. This data is in the form of 3D joint motion sequences and fed into Pose Modification which consists of a low pass filter, lower body modification, and joint position to rotation conversion. This feeds into Human Evaluations which returns user feedback back to Initial Motion Generation and passes data into Animation Conversion which consists of FBX conversion and \textit{AnimationClip} conversion. \textit{AnimationClips} are fed into the Unity Simulation Environment along with the user choice of real world background images or simulated background images in Background Generation. Unity Simulation environment outputs Pipeline Outputs consisting of synthetic images and annotations which inputted into model boosting.}
  \label{fig:DataPipeline}
\end{teaserfigure}

\maketitle

\section{Introduction}

\begin{figure*}
    \subfigure[]{
        \includegraphics[width=.18\linewidth]{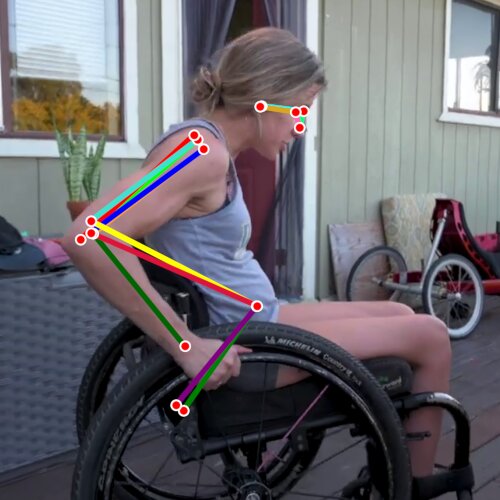}
        \label{subfig:BadImageNet1}

        \Description{A side-view of a wheelchair user pushing. Both her arms are on the wheels.}
    }
    \subfigure[]{
        \includegraphics[width=.18\linewidth]{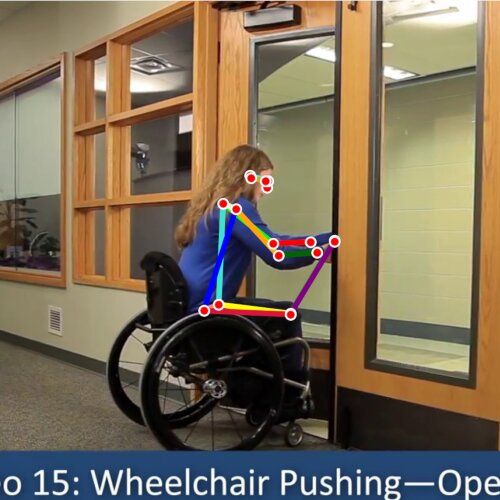}
        \label{subfig:BadImageNet2}

        \Description{A side-view of a wheelchair user opening a door. The wheelchair is directly in front of the door and both her hands are placed on the handle.}
    }
    \subfigure[]{
        \includegraphics[width=.18\linewidth]{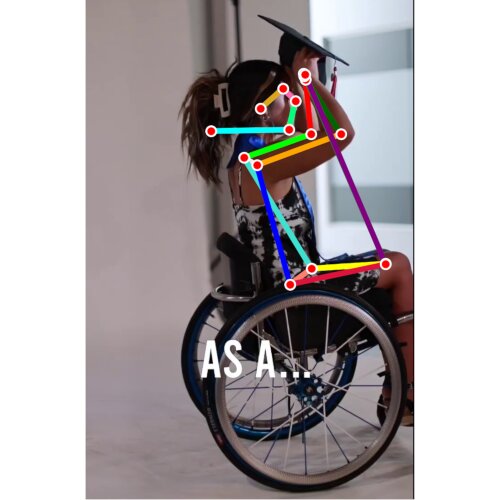}
        \label{subfig:BadImageNet3}

        \Description{A side-view of a wheelchair user putting on a graduation hat. Both their hands are holding a hang which is over her head.}
    }
    \subfigure[]{
        \includegraphics[width=.18\linewidth]{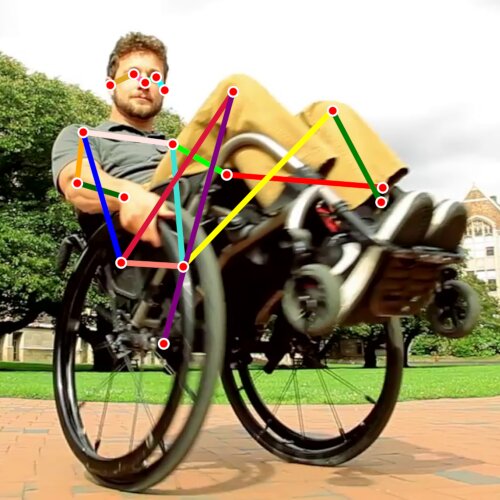}
        \label{subfig:BadImageNet4}

        \Description{A 45-degree side-front ground-up view of a wheelchair user doing a wheelie. Both hands are on the wheels of the wheelchair.}
    }
    \subfigure[]{
        \includegraphics[width=.18\linewidth]{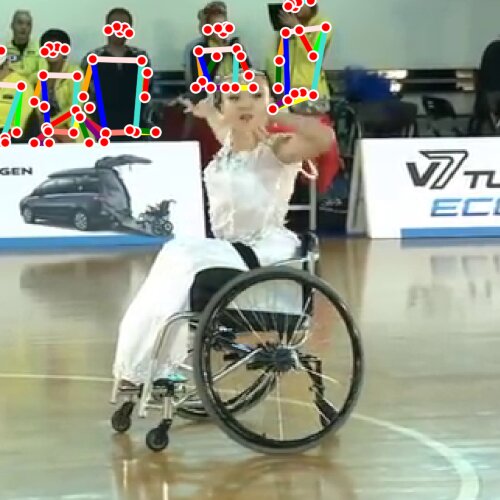}
        \label{subfig:BadImageNet5}

        \Description{A wheelchair dancer facing the camera with their wheelchair parallel to the plane of the camera. They are in front of four crowd members in a white dress.}
    }

    \caption{Examples of poor keypoint prediction performance from Detectron2 ImageNet \cite{wu2019detectron2}. 
    \Cref{subfig:BadImageNet1} Poor prediction of the legs and torso which are slightly occluded by the wheelchair. 
    \Cref{subfig:BadImageNet2} Both shins are predicted to be on the hand due to the occlusion of the wheelchair.
    \Cref{subfig:BadImageNet3} Legs are predicted to be on the upper body. 
    \Cref{subfig:BadImageNet4} The right shin is predicted to be on the wheels of the wheelchair. 
    \Cref{subfig:BadImageNet5} The wheelchair dancer is completely undetected by ImageNet.
    }
    \label{fig:BadImageNetPredictions}
\end{figure*}

The inclusiveness of AI depends on the quality and diversity of data used to train AI models. We focus on pose estimation models, which have found widespread use in health care, environmental safety, entertainment, context-aware smart environments, and more. These models are a major concern in the push for AI fairness due to the disparity in their accuracy of predicted postures between able-bodied users and users with disabilities \cite{guo2020toward,trewin2018ai,whittaker2019disability}. \changed{Focusing on human movement, Olugbade et al. \cite{Olugbade2022HumanMovementDatasets} surveyed 704 open datasets and found a major lack of diversity. The authors found no datasets that included people with disabilities performing sports, engaging in artistic expressions, or simply performing everyday tasks. We suspect that the lack of diversity has contributed to biases and poor performance on users with disabilities in many popular AI models trained on common human movement datasets like Detectron2 ImageNet \cite{wu2019detectron2}.} This is especially apparent in users who use mobility-assistive technologies (\Cref{fig:BadImageNetPredictions}). The lack of disability representation in training data can be directly attributed to poor accessibility in the data collection process for people with disabilities \cite{park2021designing, guo2020toward}. People with disabilities often must overcome more challenges in data collection during their commute and communications in the recruitment and participation process. One such example and the focus of our work are wheelchair users, who may not be able to navigate through motion capture rigs easily. Additionally, certain poses that able-bodied users could easily perform might be difficult or even dangerous for people with motor impairments. To improve disability representation in training data in the push for inclusive AI, we must make data collection equitable across people with \textit{all} levels of capabilities.

Prior work has combated this issue by proposing guidelines in the design of studies \cite{bennett2019promise} and online data collection systems that could be more accessible to people with disabilities \cite{park2021designing}. We propose an alternative solution to the data collection problem: synthetic data. Like how synthesized materials could help preserve scarce natural resources -- synthetic rubbers were developed in part because of concerns about the availability of natural rubber -- synthetic data could be valuable in supplementing insufficient data collection from people with disabilities. While traditional motion capture procedures lie on a spectrum of difficulty dictated by the motion of the user and exacerbated by the innate difficulties of data collection with users with assistive technologies, synthetic data offers a more accessible alternative where different actions, settings, and assistive technologies, ranging from cooking at home to performing back-flips in the forest, are equal in the difficulty of implementation and feasibility. 

\changed{In this research, we present a novel data synthesis pipeline which leverages motion generation models to simulate highly customizable image data of wheelchair users. Our approach includes steps for user-defined parameters, data screening, and developer feedback. This pipeline yields image data which can be used to improve the performance of AI models for wheelchair users. We evaluate this in the case of pose estimation by fine-tuning common pose estimation models, trained on common human movement datasets, with our synthetic data. Fine-tuned models are then tested on a new dataset of wheelchair users to analyze the degree of improvement from adding synthetic data to training datasets.}

Finally, as we are cautious about the problematic simulation of disabilities (e.g., blindfolded participants to simulate people without vision or with low vision), synthesized wheelchair user postures are carefully reviewed in human evaluations to avoid inadvertently exacerbating existing equitability problems our approach attempts \cite{bennett2019promise}. Our goal is not to exclude wheelchair users from AI training, but rather present a data collection solution that enables them to shepherd the synthesis of data. In doing so, this research leverages data as an intuitive way for wheelchair users to impact AI training, through which we hope to produce more fair and inclusive AI models.

Through our work, we aim to answer the following research questions and related sub-questions in the context of wheelchair users and pose estimation problems:

\begin{enumerate}
    \item \textbf{RQ1:} How to \textbf{\textit{effectively generate synthetic data}}?
    \begin{itemize}
        \item \textbf{RQ1.1:} How can we model wheelchair users?
        \item \textbf{RQ1.2:} What are the controlling parameters in synthetic data generation?
    \end{itemize}
    \item \textbf{RQ2:} What are the \textbf{\textit{efficacies of synthetic data}}?
    \begin{itemize}
        \item \textbf{RQ2.1:} How do individual parameters of synthetic data generation affect pose estimation performance?
        \item \textbf{RQ2.2:} What are the benefits and drawbacks of using synthetic data?
    \end{itemize}
\end{enumerate}

To summarize, our contribution is three-fold:

\begin{itemize}
    \item adoption of data synthesis to improve inclusion of AI.
    \item a custom data-synthesis pipeline for pose tracking with improved performance for wheelchair users.
    \item investigations of the efficacy of the overall approach.
\end{itemize}

\section{Related Work}
\subsection{Pose Estimation}
Pose estimation plays a key role in fields like the animation and video industry \cite{sutil2015motion, KinestheticIndexVideo}. Developments in deep learning have enabled users to circumvent the need for cumbersome marker suits in traditional motion capture and directly generate human postures from camera outputs \cite{xie2022physicsbased, luvizon2023sceneaware, zheng2023deep, DeepMotionAIMotion}. Of particular note is 2D posture recognition, where the recent releases of easily accessible pipelines like MediaPipe \cite{lugaresi2019mediapipe}, OpenPose \cite{cao2019openpose}, and BlazePose \cite{bazarevsky2020blazepose} have enabled widespread access to posture recognition in a wide variety of applications including biomechanics \cite{mundtEstimatingGroundReaction2022}, autonomous driving \cite{WaypointOfficialWaymo}, sign language interpretation \cite{moryossef2020realtime}, and more. 


\subsection{Sensing for People with Limited Mobility}
Currently, over 8.5\% of the population of the world is age 65 or over. This number is projected to grow to nearly 17\% of the world's population by 2050 \cite{heInternationalPopulationReports}. Given the direct correlation between age and and mobility limitations and disabilities, this trend implies a growing need for mobility-related technologies \cite{ferrucciAgeRelatedChangeMobility2016, freibergerMobilityOlderCommunityDwelling2020, gardenerMiddleAgedMobilityLimitedPrevalence2006}. Our research focuses on the community of wheelchair users, where technologies like SpokeSense \cite{carrington2019SpokeSense} have established themselves as a part of a broader focus in research related to developing smart wheelchairs \cite{leaman2017comprehensive} which integrate different sensors, including camera, lidar, and EEG, to make wheelchairs more comfortable and safe. Other works focus on developing more accessible control systems \cite{upadhyaya2022cost} or routing systems \cite{kirkham2021using} for users with mobility impairments. Posture estimation techniques for wheelchair users can reveal a user's sitting habits, analyze their mood, and predict health risks such as pressure ulcers or lower back pain \cite{maPostureDetectionBased2017}.

\subsection{Synthetic Data for Computer Vision}
Computer vision models have traditionally been trained using large-scale human-labeled datasets such as PASCAL VOC \cite{everingham2010PascalVOC}, Microsoft COCO \cite{lin2015microsoft}, and ImageNet \cite{deng2009ImageNet}. While effective, these datasets are costly to produce, requiring large amounts of publicly available images, manpower, and time to create. Furthermore, these datasets are often static and offer little in the form of customizability to allow researchers and engineers to use data specific to their task. One solution to these problems is the use of data simulators. SYNTHIA \cite{ros2016Synthia}, Synscapes \cite{wrenninge2018synscapes}, Hypersim \cite{roberts2021hypersim}, and OpenRooms \cite{li2021openrooms} provide synthetic datasets for computer vision tasks related to object detection in different settings. Other robotics simulators including AI-2THOR \cite{kolve2022ai2thor}, NVIDIA Isaac Sim \cite{liang2018gpuaccelerated}, Mujoco \cite{todorov2012mujoco}, and iGibson \cite{shen2021igibson} offer a rich set of tools for embodied AI tasks. Other systems, like BlenderProc \cite{denninger2019blenderproc}, BlendTorch \cite{heindl2020blendtorch}, NVISII \cite{morrical2021nvisii}, and Unity Perception \cite{bartolome2022perception} prefer to instead enable the developer to generate their own data through highly configurable simulators. These tools and datasets have already demonstrated considerable success in deep learning-related training tasks \cite{he2023synthetic, anderson2022synthetic, tian2023stablerep}.

Synthetic humans provide further challenges due to the complexities of human bodies and the variations and limitations of a human's appearance and posture. Various approaches have been taken, using different types of simulators to generate labeled datasets. Examples of different approaches include deriving data from hand-crafted scenes \cite{bak2018domain}, custom 3D scenes with SMPL models \cite{yang2023synbody, pumarola20193dpeople, varol17_surreal, patel2021agora, bazavan2022hspace, purkrábek2023improving}, existing games like Grand Theft Auto V \cite{cao2020longterm, fabbri2018learning, hu2019sailvos, hu2021sailvos, cai2022playing}, and game engines \cite{ebadiPeopleSansPeopleSyntheticData2022, ebadiPSPHDRISyntheticDataset2022}. We were inspired by this line of work and extended upon the existing PeopleSansPeople (PSP) data generator with the Unity Perception package \cite{bartolome2022perception} using domain randomization principles which help AI models trained in simulated environments to effectively transfer to real-world tasks \cite{tobin2017domain}.

\subsection{Evaluating the Quality of Synthetic Data}

Despite the advantages of data synthesis, an implicit assumption of using synthetic data is that it should be sufficiently high-quality to achieve performance similar to real data. To evaluate the quality of synthetic data, researchers have explored a wide range of metrics. 
Emam et al. \cite{el2020practical} outlined three types of approaches to assess synthetic data utility in their book - workload-aware evaluations, generic assessments, and subjective assessments of data utility. Among them, workload-aware evaluations check if synthetic data replicates the performance of real data, widely used in data synthesis research \cite{varol2017learning, pishchulin2012articulated, gaidon2016virtual}. 
Generic assessments measure the utility indicators of real and synthetic data when the indicators are quantifiable and clear \cite{kaloskampisdata2019Data,howe2017synthetic} (e.g., the distance between their statistical indicators such as mean, average, and distribution).
Furthermore, subjective assessments involve real users to evaluate data realism. Some researchers investigate distinguishability, assuming highly-realistic synthetic data leans to be perceived as real \cite{snoke2018general}, similar to deploying a discriminator in algorithms \cite{tzikas2022realistic}. Other researchers design Likert-Scale questionnaires for realism, which are broadly adopted in clinical training simulation \cite{bartolome2022perception}. Another criterion is to collect user preferences between several synthetic samples to form a high-quality dataset \cite{weng2023diffusion,tevet2022human}. With a consistent and valid examination of synthetic data, researchers can obtain feedback to improve generation methods and understand how reliable synthetic data is. In our work, we not only conduct workload-aware evaluation but also involve subjective assessments by asking real users to evaluate the data realism. The user-in-the-loop process provides filter handles for more realistic datasets under various contexts and inspires key findings on how data realism affects performance.
\section{\systemname{} Dataset Synthesis}

\begin{figure}
        \includegraphics[width=1.0\linewidth]{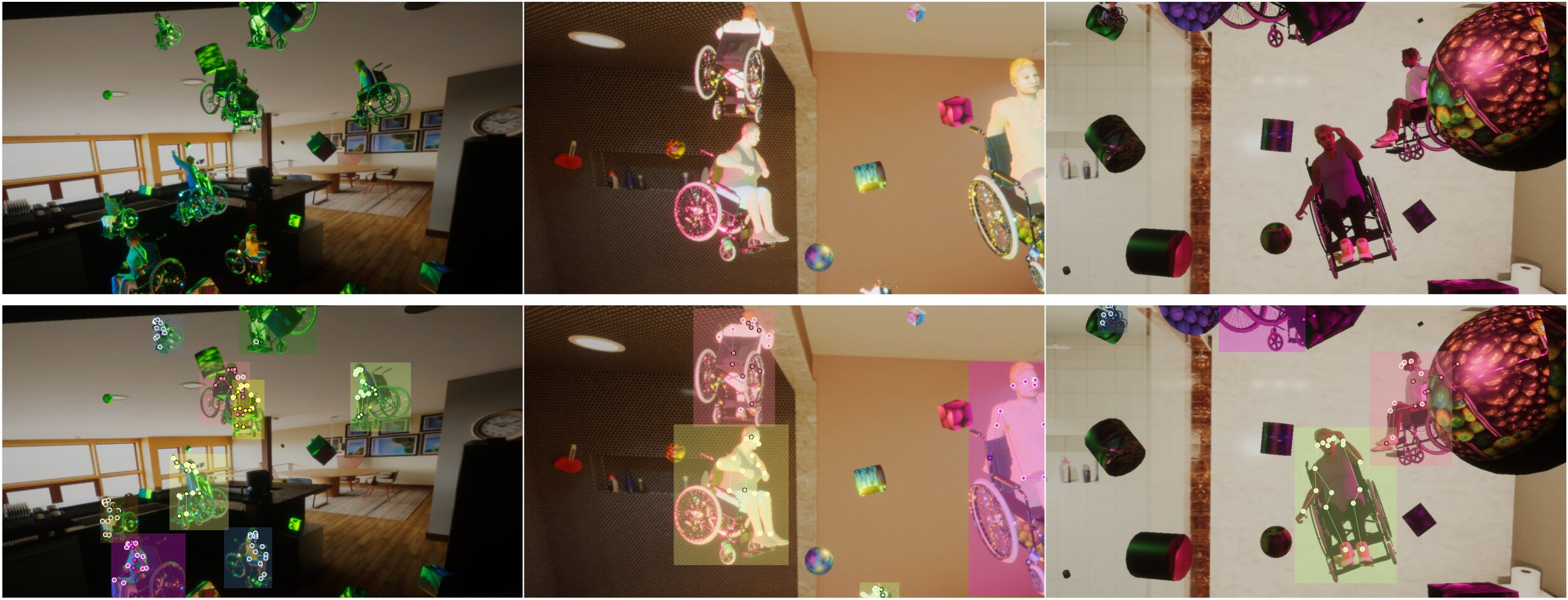}
        
        \vspace{.5cm}
        
        \includegraphics[width=1.0\linewidth]{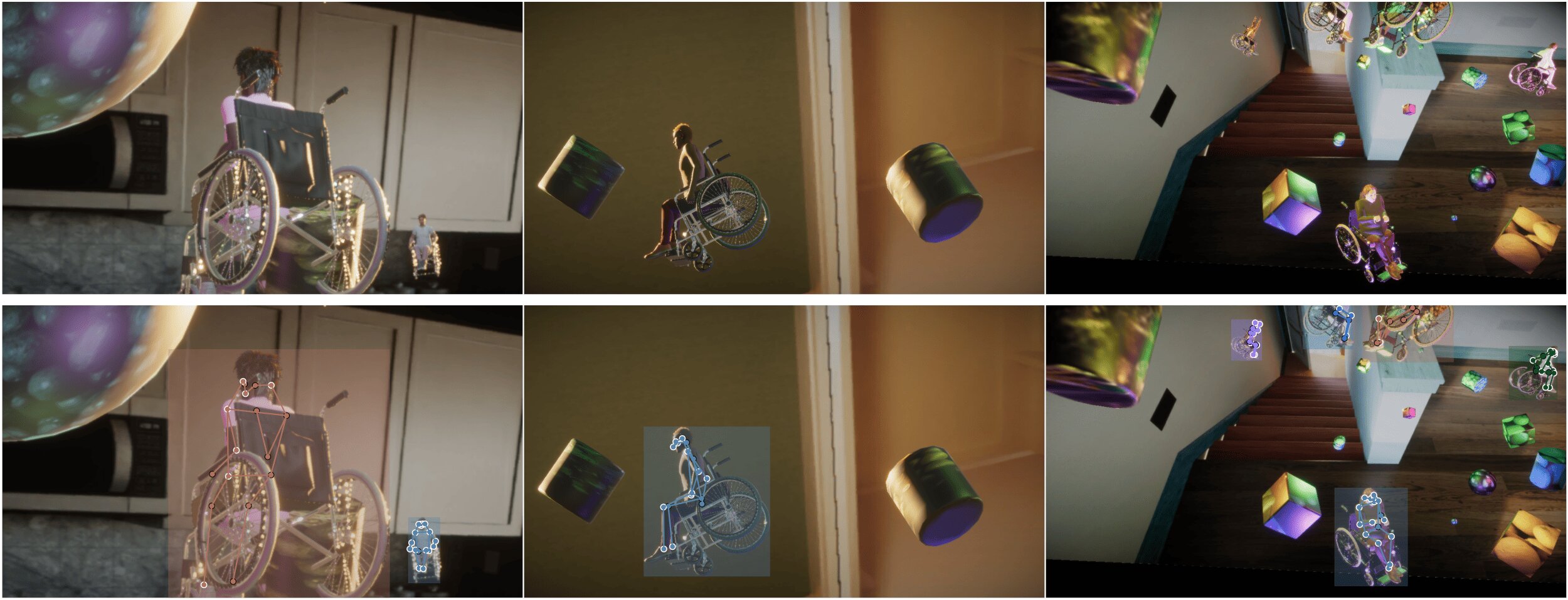}
    \caption{Example output from the full \systemname{} data generation pipeline. The top row includes the generated RGB images. The bottom row includes the generated RGB images with the keypoint and bounding box annotations superimposed on top. Keypoints outlined in black are labeled as "occluded" while keypoints outlined in white are labeled as "visible". Each image is generated with randomized backgrounds, lighting conditions, humans, postures, and occluders.}
    \label{fig:WheelPoseExampleOutput}

    \Description{Six examples of \systemname{} output. Each example features a set of different wheelchair users in front of a unique interior setting. Geometric objects with differing textures are placed around the scene. Below each image is the same image with the annotation of keypoints and bounding boxes superimposed on top.}
\end{figure}

 We address \textbf{(RQ1.1)} with our system, \systemname{}, a data synthesis framework where motion data is converted to wheelchair user animations and rigged on human models in a Unity simulation environment to generate synthetic images and annotations. We present a simple simulation environment, with human models randomly placed in front of a background as the most primitive example of synthetic data still capable of generating positive results in pose estimation (\Cref{sec:ModelPerformance}). A visualization of the overall pipeline is found in \Cref{fig:DataPipeline}. Our pipeline generates a set of datasets, each including 70,000 captured frames ($1280\times720$). Each frame is fully annotated using the COCO 17-keypoint 2D skeletal model, shown in \Cref{subfig:COCOKeypoints}. Beyond the fact that our dataset is the first fully annotated dataset of wheelchair users, the size of our dataset is comparable with existing datasets like 3DPW with 51,000 captured frames \cite{macard2018recovering} or SMPLy with 24,428 captured frames \cite{leroy2020smply}. Example synthesized data is shown in \Cref{fig:WheelPoseExampleOutput}. 

\subsection{Generating Postures}
\label{sec:GeneratingPostures}

\begin{figure}
    \includegraphics[width=1.0\linewidth]{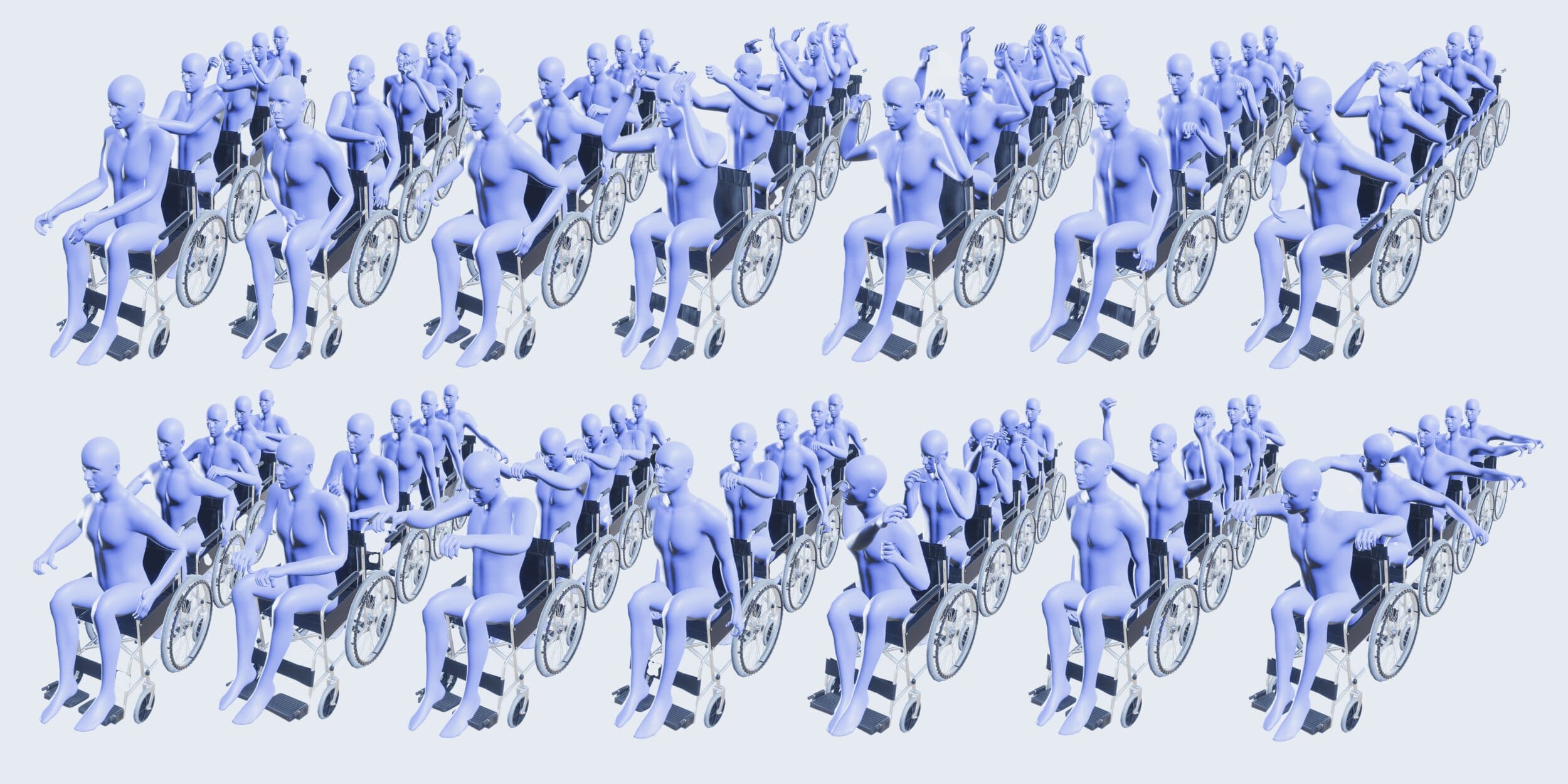}
    \caption{Example pose frames in their motion sequences resulting from our pose frame generation. Each column is an individual animation with pose frames selected in chronological order from back to front.}
    \label{fig:motionsequence}

    \Description{Models in wheelchairs performing different motions. Fourteen columns of five humans each are in a scene. Each column features different upper body movements with a large variety of ranges of motion across.}
\end{figure}

Our data synthesis pipeline begins with pose generation, where motion data is converted into animations to be used in data synthesis. We choose to use two motion data sources, HumanML3D \cite{Guo_2022_CVPR} and Text2Motion \cite{guoGeneratingDiverseNatural} in our case study. Other motion sources can be easily adapted and used within our pipeline.  \Cref{fig:motionsequence} demonstrates 14 motion sequences and their resulting postures from our pose generation, documented next. We separate posture generation from the rest of our pipeline to allow developers to iterate upon generated postures through human evaluation, regenerating and filtering data as needed.

\subsubsection{Human Skeletal Modeling} 
 
\begin{figure}

    \subfigure[\systemname{} Animation Keypoints]{
        \includegraphics[width=.45\linewidth]{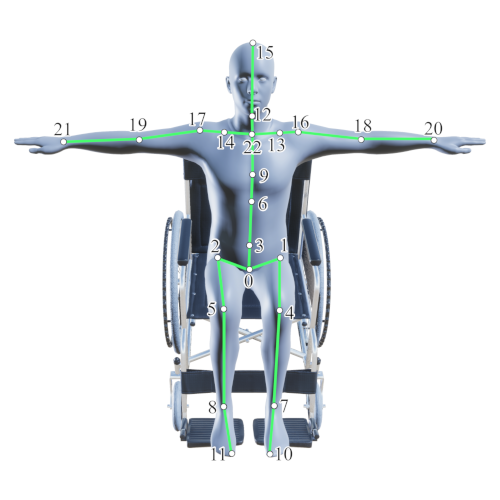}

        \label{subfig:WheelPoseAnimationKeypoints}
        
        \Description{Visualization of the \systemname{} keypoint designation. A human model sits on a wheelchair in a T-pose facing forward. Twenty-three keypoints extend across their body, going through the wrists, elbows, and shoulders to create the arms, the top of the head, the top of the neck, the neck base, the upper chest, the upper spine, the lower spine, and the middle hip to create a spine, and hips, knees, ankles, and toes to create legs.}
    }
    \subfigure[COCO Keypoints]{
        \includegraphics[width=.45\linewidth]{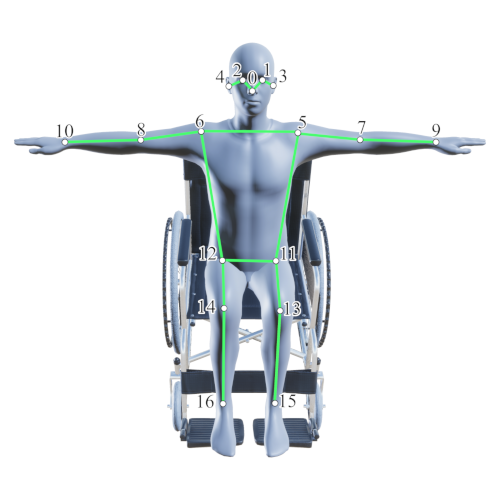}
        \label{subfig:COCOKeypoints}
        
        \Description{Visualization of the COCO keypoint designation. A human model sits on a wheelchair in a T-pose facing forward. Seventeen keypoints extend across their body, going through the wrists, elbows, and shoulders to create arms, ears, eyes, and nose to create a face, and hips, knees, and ankles to create legs.}
    }
    \caption{Keypoint mappings used in WheelPose. \Cref{subfig:WheelPoseAnimationKeypoints} WheelPose 23-keypoint animation format. Used as the input format of motion sequences before pose modification. \Cref{subfig:COCOKeypoints} COCO 17-keypoint annotation scheme. Used as the final output format of WheelPose annotations.}
    \label{fig:KeypointMappings}
\end{figure}

In order to enable a wide variety of different postures, we base the implementation of \systemname{} on a 23-keypoint skeletal model, which is easily converted from commonly available human posture datasets including COCO \cite{lin2015microsoft} and MPII \cite{andriluka14cvpr} and the output of common pose estimation algorithms including BlazePose \cite{bazarevsky2020blazepose}, MediaPipe \cite{lugaresi2019mediapipe}, and OpenPose \cite{cao2019openpose}. Detailed information on these keypoints is shown in \Cref{subfig:WheelPoseAnimationKeypoints}. \changed{We note that our current pipeline assumes users have all four limbs and acknowledge that data synthesis for wheelchair users with amputation requires efforts beyond simple ad-hoc removals of key points in our current model. Nonetheless, we interviewed two participants with limb loss, leading to insights for future work which we will discuss later in the paper (\Cref{sec:HumanEvaluation}).}

\subsubsection{Motion Sequence Generation} 
 
We use HumanML3D \cite{Guo_2022_CVPR}, a 3D human motion dataset collected from real-world motion capture in the form of 3D joint positions as an example of motion sequence generation from existing motion capture datasets. Motions with high translational movement, high lower body movement, and broken animations (e.g., jittering, limb snapping, unrealistic rotations) were filtered out from HumanML3D. Individual motions were then randomly sampled and evaluated for their uniqueness and range of motion compared to previously collected data until 100 unique motions were collected.
 
Additionally, we leverage Text2Motion \cite{guoGeneratingDiverseNatural}, a motion generation model that uses textual descriptions to generate motions, as a fully generative alternative source of human motion sequences. Text2Motion is only one example of human motion generation through textual descriptions \cite{zhang2022motiondiffuse, azadi2023makeananimation, zhang2023t2mgpt} and can be easily substituted in our data generation pipeline. Evaluators from the research team assigned textual descriptions to each of the 100 selected HumanML3D motions before inputting the descriptions into Text2Motion and generating 3 potential motions for each description. The most realistic motion was selected and evaluated on whether the motion would be possible for the evaluator to perform and the absence of any noise or artifacts from the generation process that may lead to unrealistic limb movements or positions. Our Text2Motion generation process results in a new dataset of 100 human motions that directly mirror the actions of the sampled HumanML3D motions and provide a direct comparison between synthesized and real data. Our goal in enabling the use of generative motion models like Text2Motion is to investigate the feasibility of using generative deep learning models to further simplify the data collection process for synthetic data generation and therefore improve the efficiency and overall accessibility, especially in the context of users with disabilities and assistive technologies.

In total, this process yielded 200 motion sequences from both HumanML3D and Text2Motion (i.e., each yielded 100 sequences). On average, HumanML3D motions had 60.26 (SD=42.05) unique frames per animation  and Text2Motion has 146.32 (SD=50.84) frames per animation. Since Text2Motion has no set animation length parameter, we choose to take the full animation output for each motion, leading to the discrepancy in average motion lengths, to directly compare data sourced from motion capture and deep learning models. Text2Motion outputs tend to extend and slow down the described action, leading to a longer but not necessarily more diverse animation compared to HumanML3D.

\subsubsection{Pose Modification and Conversion}
\label{sec:PoseModification}

Both HumanML3D and Text2Motion represent motions through the 3D position of joints. All motions are converted into the corresponding 23-keypoint skeletal model used by \systemname{} through a direct mapping of corresponding joints or the positional average between the surrounding joints. As is common in biomechanical analysis \cite{schrevenOptimisingFilteringParameters2015, bartlettIntroductionSportsBiomechanics2014}, a 5Hz low pass filter is then applied to the data to handle high-frequency noise generated from the motion capture or data generation process. We convert all motions from a 3D position to a joint rotation representation. In the process of this conversion, internal and external rotations -- rotations around the axis parallel to the bone not able to be described in joint position notation -- of the arms are affixed to the rotation of the parent bone. 

We modified the resulting pose sequences from the two generation methods by affixing the model's legs onto the wheelchair model with an additional rotational noise applied independently on the flexion/extension, abduction/adduction, and internal/external rotation on each of the lower body bones (i.e. three joints in total) to simulate regular lower body movements when in a wheelchair. The following steps document the procedure for generating rotational noise on one joint. This procedure was motivated by the need for smooth interpolated noise and inspired by the Poisson process. 

Given an array of joint angles $F_{orig}$ expressed in degrees with length $n$ total frames in the animation, the noise is generated by first sampling a set of frames indices $S$.

\begin{equation}
    \textstyle S = \left\{ x_0,x_1,...,x_n\ \middle\vert \begin{array}{l}
        x_0 = 0, \\
        x_i = x_{i-1}+k_i, \\
        k_i\sim\mathcal{N}(\frac{n}{4}, (\frac{n}{32})^2), \\
        x_i < n, \\
        i = 1, 2,..,n
      \end{array}\right\}
\end{equation}

A new array of joint angle noise values $N$ is then constructed. Let $f(i)$ be a function that generates the $i$-th joint angle of the animation. Given $U\sim\mathcal{U}(-10,10)$, representing a random angular noise added to frames in $S$,

\begin{equation}
    f(i) = 
    \begin{cases}
        U & i \in S \\
        \text{NaN} & else
    \end{cases} 
\end{equation}

\begin{equation}
    N = [f(i) \mid 0 \leq i < n]
\end{equation}

Linear interpolation is then applied to fill in all NaN values in $N$. The new array of joint angles $F_{new}$ is thus the element-wise sum of $F_{orig}$ and $N$. Our algorithm is motivated by the need for a simple and efficient noise algorithm that does not jitter as the user iterates through frames of the animation.

\begin{equation}
    F_{new} = [f_{orig}[i] + N[i] \mid 0 \leq i < n]
\end{equation}

All motion sequences are applied to a Blender\footnote{A free and open source 3D modeling software. More information is found at \url{https://www.blender.org/}} model before being imported into Unity and converted to Unity Perception-readable \textit{AnimationClips} files. Example outputs from our posture generation step described above can be found in \Cref{fig:motionsequence}.

\subsection{Generating Wheelchair User Models}

In order to capture synthetic images, we must rig the postures resulting from previous sections onto Unity human models.

\begin{figure}
    \includegraphics[width=1.0\linewidth]{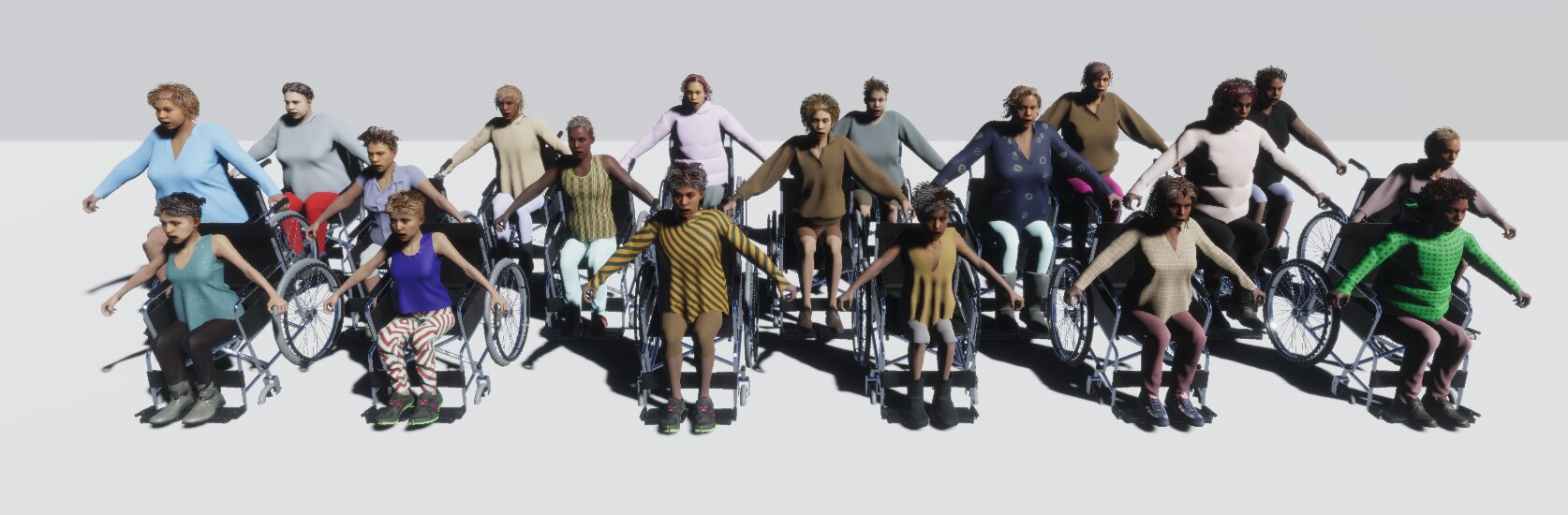}
    \caption{Example human models used in the synthesis of wheelchair user images.}
    \label{fig:humanmodel}

    \Description{Nineteen unique human models in wheelchairs. Each model has a different combination of clothing, age, ethnicity, sex, hairdo, and body size.}
\end{figure}

For human models, \systemname{} enables both the default human models provided by PeopleSansPeople \cite{ebadiPeopleSansPeopleSyntheticData2022} and randomized humans leveraging the Unity SyntheticHumans \cite{SyntheticHumansPackageUnity2023}, a Unity Perception package using domain randomization to generate unique human models from a sampling of different clothes, body types, sexes, and more. This utilization allows \systemname{} to generate unique human models that better capture the wide variety of appearances of real people. We use 8,750 unique human instances using the default SyntheticHumans configuration limited to people over the age of 10 to better reflect the general population of people in wheelchairs \cite{vignier2008WheelchairDemographics}. \Cref{fig:humanmodel} shows 19 examples of these human models.

We also enable the human models to spawn with different objects (e.g., wheelchairs, crutches, canes, walkers, etc.) in user-defined positions when placed into the environment. For the scope of this project, we focused on wheelchair users and used a realistic wheelchair model, sourced from the Unity Asset Store, scaled by the size of the human model. Finally, the posture of each human model is randomly sampled from the \textit{AnimationClips} generated in \Cref{sec:GeneratingPostures}. 

\subsection{Generating the Simulation Environment}

We address \textbf{(RQ1.2)} by developing a highly configurable simulation environment. We use the PeopleSansPeople (PSP) \cite{ebadiPeopleSansPeopleSyntheticData2022} base template and its related extension, PSP-HDRI \cite{ebadiPSPHDRISyntheticDataset2022}, as our baseline data generator built in the Unity\footnote{More information found at \url{https://unity.com/}} game engine through the High Definition Render Pipeline (HDRP). PSP is a parametric human image generator that contains a fully developed simulation environment including rigged human models, parameterized lighting and camera systems, occluders, synthetic RGB image outputs, and ground truth annotations. PSP is built on the idea of \emph{domain randomization} \cite{tobin2017domain} where different aspects of a simulation environment are independently randomized to diversify the generated synthetic data, exposing models to a wider array of different environments during training and improving testing accuracy \cite{tremblay2018training,valtchevDomainRandomizationNeural2021}. All domain randomization is implemented through the "randomizer" paradigm designed in the Unity Perception package \cite{borkman2021unity}, a Unity toolkit to generate synthetic data for computer vision training. Within each scene, individual randomizers are assigned to a specific parameter (i.e. lighting, occluder positions, human poses, etc.) and independently sample parameter values from a uniform distribution. PSP was then updated to Unity 2021.3 and Unity Perception 1.0.0\footnote{First official release of Unity Perception} which enabled more annotations, more extendable randomizers, and flexibility with other Unity packages. We then added a new background image parameter and its related randomizer for the sampling of user-defined images to be used as a backdrop in each scene. We use this parameter to enable three different sets of background images: PSP default textures, 100 background images randomly sampled from the BG-20k background image dataset \cite{li2022bridging, li2021privacypreserving, li2021deep}, and 100 generative images from Unity SynthHomes \cite{SyntHomes}, a dataset generator for photorealistic home interiors. \Cref{tab:EnvironmentParameters} outlines the statistical distributions of the environment parameters used.

\subsubsection{Camera Configuration and Keypoint Annotations}

A main camera in the Unity scene is used as the primary capture source of all images and annotations. The Unity Perception \emph{Perception Camera} is used to simulate realistic camera features including focal length and field of view (FOV) and to capture all annotations. The position, rotation, focal length, and FOV of the main camera are set through a series of randomizers with default parameters found in \Cref{tab:EnvironmentParameters}. The main camera captures RGB, depth, surface normal, and instance segmentation images in $1280\times720$ for each frame of captured data. Default Unity Perception annotation labelers are placed on the main camera to capture 2D/3D bounding boxes for each human model, object counts, rendered object metadata, semantic segmentation, 2D/3D keypoint locations in COCO format, and percent of human model occluded. Out-of-view and fully occluded human instances are automatically ignored in annotation capture, recording only data on human instances within direct view of the main camera.

\subsection{Assembling \systemname{} Datasets}

\begin{figure*}
    \includegraphics[width=.98\linewidth]{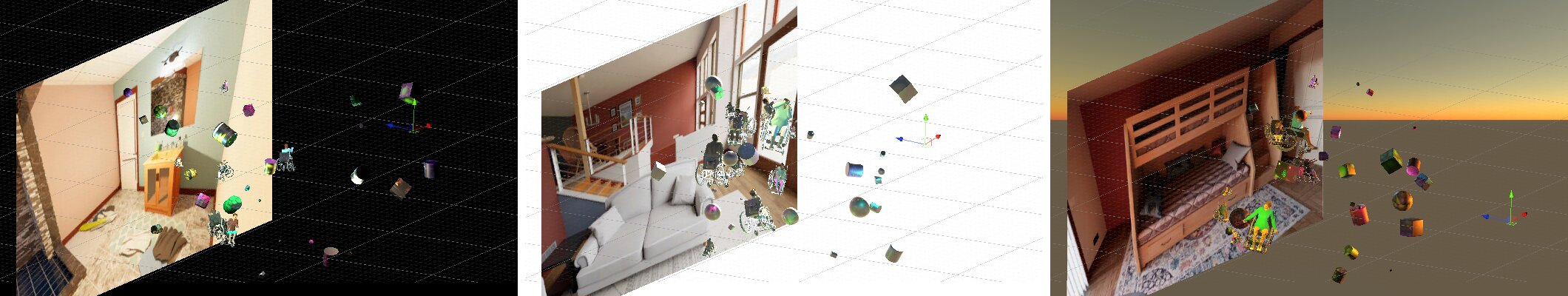}
    \caption{Examples of a scene being generated. Notice the random placement of wheelchair users, different occluder objects, different lighting conditions, and various SynthHome backgrounds. The red, green, and blue arrows represent the camera coordinates which we used to insert randomization in forms of Cartesian translations and Euler rotations.}
    \label{fig:ExampleScene}
    
    \Description{Three examples of scene generation in \systemname{}. Each example features a third-person example of wheelchair users in front of a backdrop with occluders randomly suspended in the air. Arrows indicating the camera are all placed in different locations and rotations for each image.}
\end{figure*}

 Overall, our pipeline yields 70,000 images for each generated dataset. All data was generated in a Unity 2021.3 project configured with parameters preset to the values listed in \Cref{tab:EnvironmentParameters}. We ran our data synthesis pipeline on a PC with a 4.2GHz 6-Core/12-Thread AMD Ryzen R5 3600, NVIDIA GTX 1070 8GB VRAM, and 32 GB 3600MHz DDR4 memory for an average generation time of $\approx$ 1 hour and 45 minutes for 10,000 images -- which translates to 12 hours and 15 minutes for each dataset. This time includes all steps of the generation process including motion generation, parameter randomization, data capture, label creation, and writing to disk. Examples of generated synthetic images are shown in \Cref{fig:ExampleScene}. We open-source our data synthesis pipeline including pose modification and the full configurable Unity 2021.3 project for data generation in \url{https://github.com/hilab-open-source/wheelpose}.

\section{Evaluation of \systemname{}}
\label{sec:Evaluation}

To answer \textbf{(RQ2)} on the benefits and drawbacks of synthetic data, we evaluate the \systemname{} pipeline and generated data through three specific methods: 1) human evaluation on realism, 2) statistical analysis of innate dataset characteristics, and 3) evaluation of our generated datasets' effects on AI model performance. We document the results of these methods in the following sections. 

\subsection{Human Evaluation}
\label{sec:HumanEvaluation}
We involved real wheelchair users in the loop to evaluate the realism of our synthetic data. In our study, ``realism'' manifests as \textbf{\textit{ease}} and \textbf{\textit{frequency}}. 
\changed{We sent out online surveys and strictly verified users' eligibility and authenticity manually to prevent scammers. We recruited 13 daily wheelchair users (5 F, 8 M), with ages ranging from 26 to 56 (M=32, SD=9.6), as shown in \Cref{tab:HumanEvalDemographics}.} \changed{A key limitation of this research is the limited diversity among the wheelchair user participants. All participants were individuals with spinal cord injuries (SCI), a specific condition that has distinct movement patterns. This represents only a small subsection of the broader wheelchair user population. Despite our efforts to diversify the participant pool by including participants with different levels of SCI, we acknowledge that our participant population is not sufficiently representative to draw statistical insights for wheelchair users with different conditions or bodies than those in this study (e.g., muscular dystrophy, amputations, dwarfism, spinal deformities).}

\begin{table*}
\caption{Demographics of participants (P1-P13) in human evaluation.}
\label{tab:HumanEvalDemographics}
\resizebox{\linewidth}{!}{
\begin{tabular}{@{}lllllp{6cm}p{4cm}@{}}
\toprule
\textbf{ID} &
\textbf{Age} &
\textbf{Gender} &
\textbf{Occupation} &
\textbf{SCI Level} &
\textbf{Exercise Routines} &
\textbf{Full Mobility of Arms, Shoulders, and Hands} \\ \midrule
P1 &
  56 &
  M &
  Professor &
  T-12/L-1 &
  No &
  Yes \\
P2 &
  29 &
  F &
  Home-maker &
  T-3 &
  Weight lifting, Strolling, and Stretches &
  Yes \\
P3 &
  40 &
  M &
  Entrepreneur &
  T-12 &
  Swimming, Cycling, and Gym &
  Yes \\
P4 &
  40 &
  F &
  Self-employed &
  C-5 &
  No &
  No (DASH Score 79.2/100 \cite{hudak1996development}) \\
P5 &
  36 &
  M &
  Customer Relations &
  C-5 &
  Stretching and Strength training &
  No \\
P6 &
  26 &
  M &
  Software Developer &
  Lumbar spinal stenosis &
  Wheelchair walking &
  Yes \\ 
P7 &
  51 &
  M &
  Proctor/Graphic Designer &
  T-12; L1 &
  Gym workouts periodically, Wheelchair Basketball, Wheelchair Tennis, and Pushing long distances &
  Yes \\ 
P8 &
  26 &
  M &
  Librarian &
  C7 &
  No &
  No (DASH Score 40.0/100 \cite{hudak1996development}) \\ 
P9 &
  27 &
  F &
  Remote Computer Programmer &
  L5 &
  Arm training using a band &
  Yes \\ 
P10 &
  27 &
  F &
  Teacher &
  Lumbar SCI &
  Aquatic therapy &
  Yes \\ 
P11 &
  32 &
  F &
  Receptionist &
  Thoracic SCI &
  Aerobic exercise &
  Yes \\ 
P12 &
  34 &
  M &
  Marketing Manager &
  Sacral SCI &
  No &
  Yes \\
P13 &
  29 &
  M &
  Freelancer &
  Lumbar SCI &
  Water exercise &
  Yes \\
\bottomrule
\end{tabular}}
\Description{This table shows demographic information about participants in the human evaluation. The header includes ID, Age, Gender, Occupation, SCI Level, Exercise Routines, and Full Upper Limb Mobility.}
\end{table*}

\subsubsection{Procedure} 

\begin{figure*}
    \includegraphics[width=0.7\linewidth]{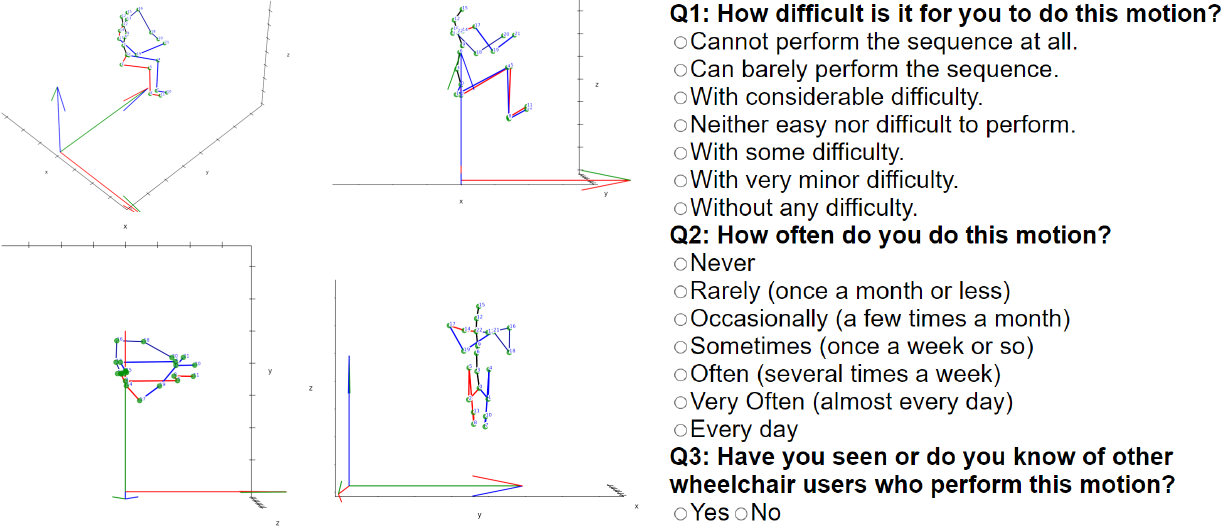}
    \caption{Screenshot of the human evaluation interface. Note that the four subplots on the left are supposed to show an animation of a human skeleton performing a motion sequence in loops from different perspectives (45-degree oblique top view, top view, side view, and front view). Participants were asked to observe an animation and give Likert-Scale scores and a binary response before moving on to evaluation of the next motion sequence.}
    \label{fig:HILAnimationClipExample}

    \Description{Example of human evaluation interface. Four different views of a skeletal model performing one of the selected motions are on the left. On the right, a set of questions on motion difficulty, frequency, and whether they have seen the motion before with multiple choice answers are displayed.}
\end{figure*}

Participation was conducted entirely online, allowing users to contribute at their convenience. Users first answered a questionnaire consisting of demographic information and mobility capability before then  evaluating two groups of synthetic motions - HumanML3D motions and Text2Motion motions, using a browser-based user interface (\Cref{fig:HILAnimationClipExample}). The motions were presented as animation GIFs of a human skeleton performing a certain movement. We chose to use skeletons rather than a more photorealistic model as an embodiment technique facilitating users to think of these skeletons as tracked motion of their bodies. Related work \cite{fribourg2020avatar} has shown that a dummy avatar provides stronger senses of embodiment comparable to non-personalized realistic avatars. Our uses of the skeletal model aim to shift the focus towards motion, steering attention away from superficial cosmetic details. Users observed one animation clip at a time, simultaneously from four perspectives (45-degree oblique top view, top view, side view, and front view). To navigate through motion clips, users click the previous/next buttons or press the arrow keys. Buttons were designed for participants to select scores for two Likert-Scale questions.

For each motion, users answered three questions:

\begin{itemize}
    \item \textbf{Q1:} ``How difficult is it for you to do this motion?'' - by rating from 1 (Cannot perform the sequence at all) to 7 (Without any difficulty). 
    \item \textbf{Q2:} ``How often do you do this motion?'' - by rating from 1 (Never) to 7 (Everyday).
    \item \textbf{Q3:} ``Have you seen or do you know of other wheelchair users who perform this motion?'' (Yes/No).
\end{itemize}

We used all 100 motions in both HumanML3D and Text2Motion converted motion sets. After finishing the last animation GIF, users proceeded to the other motion group. The question set is consistent for both groups. The order of groups was random. Three users evaluated HumanML3D first, while the others saw Text2Motion first. 

After all animations were scored, we followed up with participants via email to better understand the following: 

\begin{enumerate}
    \item What are the criteria used in your evaluation of a motion's difficulty?
    \item What determines the "frequency" of performing a motion?
    \item What motion do most wheelchair users often perform and what motion do you frequently perform, that did not show up in our dataset?
\end{enumerate}

Participants were paid \$20 per hour as compensation for their time. As the study did not enforce a time limit and was purely online, users could take a brief break whenever they wanted as long as the questionnaire remained open. 
\changed{Excluding breaks, the study took 1.78 hours on average.} The study was evaluated and approved by the Institutional Review Board (IRB) at UCLA.

\subsubsection{Data Analysis}

\begin{figure*}
    \includegraphics[width=0.9\linewidth]{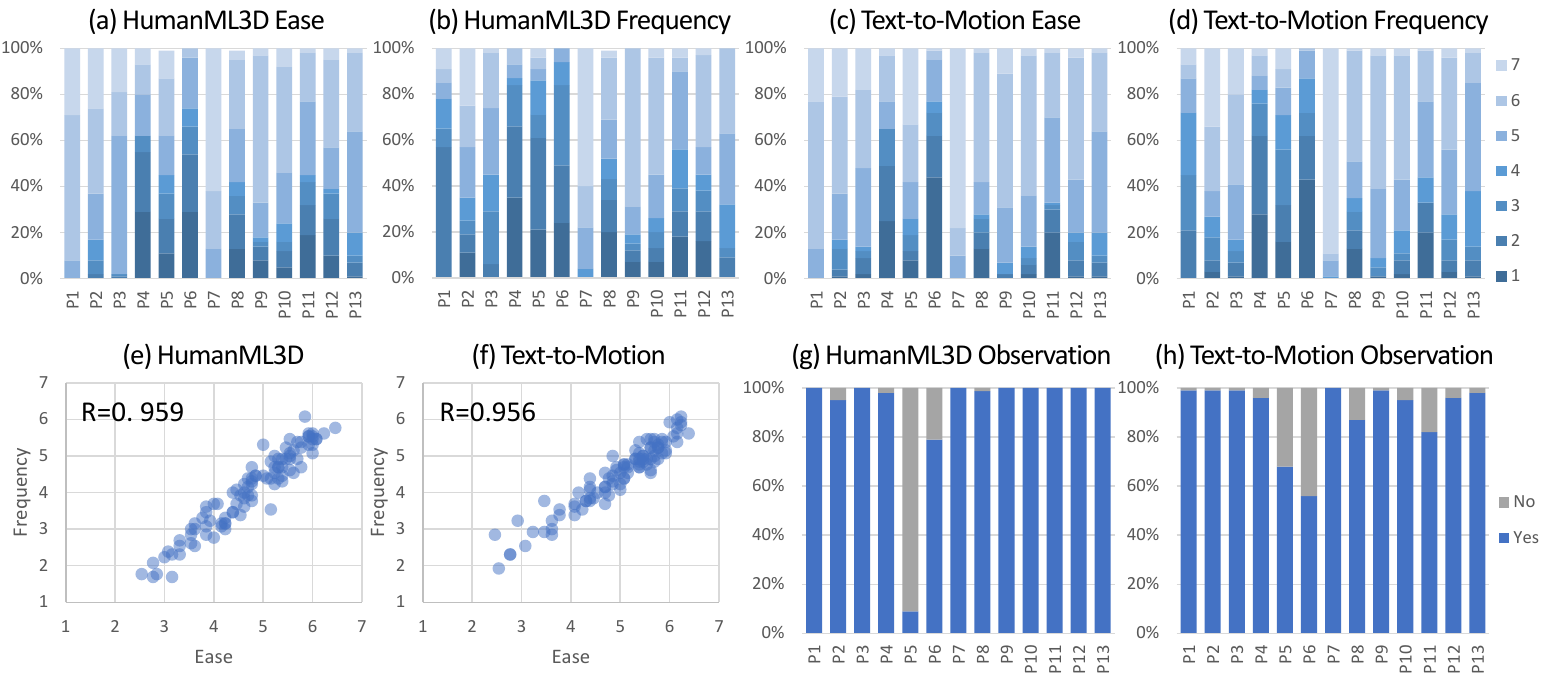}
    \caption{From left to right, Figure (a)(b)(c)(d) are 100\% stacked bar charts showing the motion distribution in HumanML3D or Text2Motion, with (a) and (b) depicting the perceived ease and frequency of HumanML3D, while (c) and (d) depicting Text2Motion. In (a) and (c), a score of 1 denotes \textit{Cannot perform the sequence at all} while a score of 7 denotes \textit{Can perform without difficulty}. In (b) and (d), 1 means \textit{Never}, and 7 means \textit{Everyday}. The two scatter plots, Figure (e) and (f), demonstrate the strong correlation between ease and frequency in both motion groups. Figures (g) and (h) depict stacked bar graphs showing whether participants have seen or known of a wheelchair user who has performed this motion.}
    \label{fig:HumanEvalStat}

    \Description{Four stacked bar graphs with percent of animations from 0\% to 100\% on the Y axis and the participants on the X axis. Each bar graph has responses 1 - 7 stacked to add up to 100\%. Two more scatterplots with frequency on the Y axis from 1 to 7 and ease on the X axis from 1 to 7. Both scatter plots have strong linear correlations. Two final stacked bar graphs with Yes or No frequency on the Y axis and the participant on the X axis. A majority of participants answered yes they have seen most motion. }
\end{figure*}

User responses were Likert-Scale scores (1-7 for Q1 and Q2) and binary responses (yes/no for Q3). We first visualized the distribution of data across users (\Cref{fig:HumanEvalStat}). Afterward, we separately ranked the motions based on average ease and frequency scores. We also calculated the correlation between difficulty and frequency. 
\changed{From follow-up emails, we collected their comments and performed a thematic analysis of their perception of metrics, and the validity of our dataset. The initial codes were the summary of their rating reasons, which were later merged and discussed.}
In the end, we ranked Text2Motion motions by the total frequency and difficulty score and identified the bottom 10\% for later use in the model performance evaluation. 

\subsubsection{Results and Findings} 

User perception of the generated dataset (i.e., perceived realism) is a strong indicator of the efficacy of our data generation pipeline in that efficient data generation should only yield data that wheelchair users perceive as being realistic. Investigating deeper causes for the perceived realism of our generated data also inevitably led to insights about wheelchair users. We summarize the findings of our user evaluation in the rest of this section.

\noindpar{Participants found our synthetic motion sequences realistic.} \changed{HumanML3D motions received an average ease score of 4.742 (SD=0.943) and an average frequency score of 4.079 (SD=1.085) across the dataset. Text2Motion motions performed comparably, receiving an average ease score of 4.977 (SD=0.915) and an average frequency score of 4.507 (SD=0.889). }
Regarding Q3, for each motion we presented, our participants had seen or knew of other wheelchair users who performed that motion. Specifically, as \Cref{fig:HumanEvalStat}(g)(h) shows, most participants have seen most motions performed by other wheelchair users. 

\changed{Regarding the representativeness of datasets, most participants said they had seen all of the common motions (e.g., ``rolling forward'', ``driving'', ``writing'', ``drinking'', ``cooking'') they knew about in our datasets. We suspect two factors that account for the outcomes observed in these participants: 1) lack of contextual cues made it challenging for them to recall specific motions, and 2) our embodiment technique facilitated their use of imagination that bridged the gaps between the motions we demonstrated and those they executed in daily tasks. Nevertheless, some other participants commented on popular but missing motions, including ``wearing some lower body clothes'' (P10), ``clapping hands'' (P11), ``typing on a keyboard'' (P13). The rest of the users were unsure about their recollection of the datasets and used general phrasings e.g., ``seen almost all''. Our datasets were not comprehensive enough to cover all common motions. We believe user feedback is the key to improving data completeness and representativeness.}

\noindpar{Ease (Q1) and frequency (Q2) of performing a motion vary across participants.} From \Cref{fig:HumanEvalStat}(a) to \Cref{fig:HumanEvalStat}(d), we observe diversity of perception on ease and frequency across individuals. The horizontal axis is participant ID, the vertical axis is the percentage of motions in the whole group. The sequential color palette depicts the ease of doing a motion from 1 to 7, with 7 being very easy. For example, P4 and P5, who had a higher injury position, rated more motions into the difficult pool. The takeaway is that the ease and frequency of performing a motion are highly personal. A realistic motion dataset should include motions across the entire spectrum while eliminating motions that no wheelchair users will ever perceive as being easy/frequent to perform. However, the result of this user evaluation would change as the size of the participant group grows, which lowers the likelihood of unrealistic motions. Nonetheless, we argue that a realistic dataset should include motions that most wheelchair users would perceive as easy and frequent, to ensure the data quality and avoid introducing new biases, while addressing existing issues in the inclusiveness of data collection. Our human evaluation serves as a reference for conducting user assessments of data quality in motion synthesis.

\noindpar{There was a strong correlation between ease and frequency.} The last two scatter plots (\Cref{fig:HumanEvalStat}(e)(f)) showcase a strong positive correlation between ease and frequency. In other words, less difficulty was associated with a higher frequency of usage. 
\changed{The Pearson correlation coefficients \cite{schober2018correlation} are respectively 0.956 and 0.959 for Text2Motion and HumanML3D. }
This result was expected for motions that are difficult to perform for wheelchair users being less frequently performed because they are often circumvented by alternative motions, which was confirmed by comments from participants later on.

\noindpar{Factors on ease: range of motion, pain, balance, and tiredness.} 
\changed{Participants evaluated how easy a motion was with several factors. 7 out of 13 people highlighted their range of motion using keywords ``\textit{range of motion}'', ``\textit{being paralyzed}'', ``\textit{based on my abilities}'', ``\textit{considering the angles}.'' Six participants mentioned that pain affected their decisions. }
Regarding tiredness, P2 explained ``\textit{Since I have no core muscles, being paralyzed from the nipple line down, a lot I can do, but can’t do for long}''. Besides, P2, P4, and P5 emphasized balance. For example, P2 commented ``\textit{My left side is a little higher on my injury, so I struggle a lot with that side, or even staying upright when both hands are in use.}'' Their perception of ease was reflected in Likert-scale scores in \Cref{fig:HumanEvalStat} (a)(c).

\noindpar{Factors on frequency: utility and ease.}
How frequently one would make a motion depended on both utility and ease. This explained why the correlation coefficient was strong but not definite (e.g., R=0.99). 
An easy motion did not necessarily lead to frequent usage. Participants determined frequency mainly based on their routine. P1 commented, ``\textit{Cooking, cleaning, driving, shopping... based on where/when/how I might use the motion, I decided how often I might actually use the motion}.'' 
\changed{Along with P1, six more users recalled their daily routine when rating for the frequency. P3 and P8 also referred to exercises/therapies to define the frequency of each motion. }

On the other hand, when difficulty and utility had conflicts, participants were experts in circumventing them with alternative motions if possible. P4 explained, ``\textit{I'm limited in my arm functions with putting hands or arms over my head and behind my back... so I use straps on shoulders and high back on my wheelchair}.'' P5 also commented on how he used elbows and hands to support himself without abdominal muscles when needed, ``\textit{As a tetraplegia, I don't really have abs. So, I'm kinda 'crawling' to get back up. Propping myself up with elbows on knees and such.}'' 

\noindpar{Sense of embodiment.}
Results show that our uses of the skeletal model successfully facilitated the sense of embodiment -- participants could think of the human model as their own body, and imagine themselves performing those motions when evaluating the ease and frequency. P1 commented, ``\textit{I mimicked the motions and tried to decide when/how I would use the motion}''. Similarly, another comment said, ``\textit{I picture myself doing certain tasks throughout the day, or motions I use for exercise}.''
On this note of sense of embodiment facilitation, we also chose not to attach text descriptions to animation clips, but encouraged users' imagination by allowing them to interpret the visuals on their own.

\changed{We expanded our study and recruited two new amputee participants (U1 and U2, not from P1-P13). They (1F, 1M) both had an artificial leg and used wheelchairs every day. We asked open-ended questions about their expectation of pose estimation, including but not limited to an amputated skeleton, the integration of artificial limbs, and the desired features/interfaces for pose estimation. To increase the sense of embodiment, U1 and U2 both proposed using an amputated model that accurately represented their body. For example, U1 commented, ``To be accurate I think the skeleton should be amputated to maintain accurate pose estimations.'' Similarly, U2 said the model ``should be able to be customized to meet different users with different amputations.'' 
However, expectations changed when artificial limbs came in. U2 talked about our research's reflection of his artificial limb, and said ``The skeleton should adapt to work together with the artificial part that is added.'' Meanwhile, U1 insisted the integration of artificial limbs depended on users, saying ``I think should allow users to access settings and have a choice of their own.'' 
As for the feature/interface, both U1 and U2 mentioned a settings panel to annotate where a person was amputated, and U1 further suggested an option to turn on/off the integration of artificial limbs.}

\begin{figure}[t]
    \centering
        \subfigure[Number of Bounding Boxes per Image]{
            \includegraphics[width=.46\linewidth]{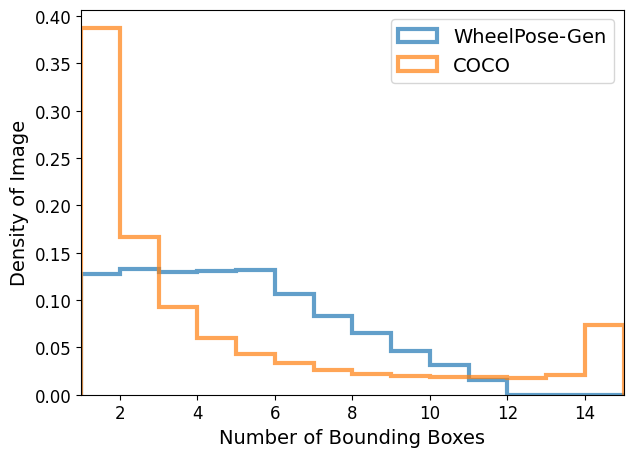}
            \label{subfig:NBBoxes}

            \Description{Histogram showing the density of images from 0 to 0.4 on the Y axis against the number of bounding boxes from 0 to 15 on the X axis. Two lines are shown. \systemname{}-Gen was more evenly distributed from 0-12 bounding boxes at around 0.13 density while COCO peaked at 1 bounding box at 0.38 density and quickly descended to 0.04 at 10 bounding boxes before slightly jumping to 0.10 at 15 bounding boxes.}
        }
        \subfigure[Relative Size of Bounding Box]{
            \includegraphics[width=.46\linewidth]{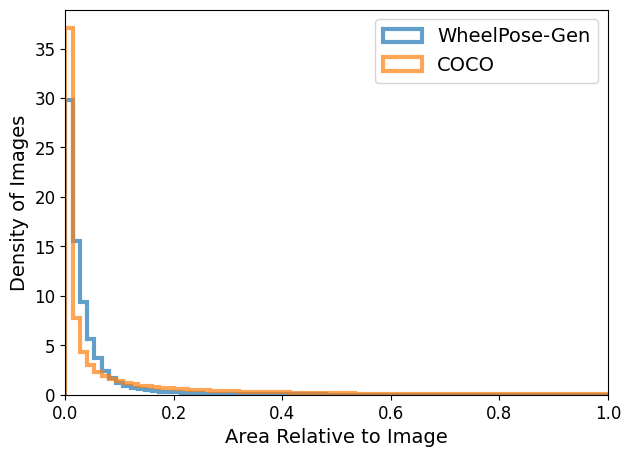}
            \label{subfig:BBoxArea}

            \Description{Histogram showing the density of images from 0 to 35 on the Y axis and the area of a bounding box relative to the image from 0 to 1 on the X axis. Two lines are shown. Both \systemname{}-Gen and COCO follow a similar exponential decay-shaped distribution, starting at a density of 30 and 38 respectively, and dropping to nearly 0 when the area relative to the image is greater than 0.2. }
        }
        \subfigure[Bounding Box Spatial Location Heatmap]{
        \includegraphics[width=.98\linewidth]{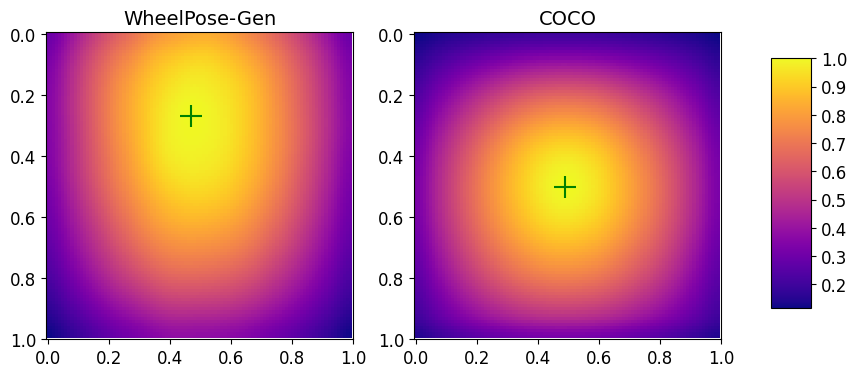}
        \label{subfig:BBoxHeatmap}

        \Description{Two heatmaps showing the distribution of bounding boxes. \systemname{} has a more oblong distribution stretching upwards into the top edge of the heatmap. COCO has a very circular distribution centered right in the middle of the heatmap with little to no distribution at the edges of the heatmap.}
        }
    \caption{Bounding Box Statistics. All COCO statistics are computed for images that contain at least one person instance in COCO.
    \Cref{subfig:NBBoxes} Number of bounding boxes per image. Images with no human bounding boxes are not counted as COCO is not a purely human dataset. 
    \Cref{subfig:BBoxArea} Relative area of each bounding box compared to the image area. Relative area is computed through $\sqrt{\text{bounding box area}/\text{image area}}$. 
    \Cref{subfig:BBoxHeatmap} Heatmap of bounding box location scaled by image size.  
    The color of each pixel maps to the likelihood of that corresponding coordinate in an image being bounded by bounding boxes. The peak location of the heatmap is marked with a green $+$.}
    \label{fig:BBoxGeneral}
\end{figure}

\subsection{Statistical Analyses}

We performed a set of statistical analyses to examine how the diversity and size of our dataset compare to the full COCO 2017 persons dataset (training and validation) \cite{lin2015microsoft}, selected as our benchmark for its ubiquity in other 2D human pose estimation related research \cite{cao2017realtime,sun2019deep,xu2023vitpose}. Greater diversity and size in training datasets has been shown to improve machine learning performance \cite{Gong2019diversity} and is a common solution to prevent overfitting. We consider the following categories in our analyses: high-level dataset features, bounding box location, size, and number in generated images, keypoint number and occlusion per image and instance, diversity of human poses, and camera placement. For all subsequent dataset statistics, we used the \systemname{}-Gen dataset, which was generated with no real-world data using SyntheticHumans models, Text2Motion animations, and SynthHome background images. The other datasets from \systemname{} shared similar statistics and were skipped in this validation to avoid redundancy.

\subsubsection{High-Level Dataset Features}

In total, our dataset has 70,000 images with bounding boxes and keypoint annotations. We chose to generate 70,000 images to mimic the size of the COCO dataset. For reference, a recent effort in synthetic data has proven that a PSP dataset of as little as 49,000 images was found to have meaningful pose estimation improvements in Detectron2 fine-tuning \cite{ebadiPeopleSansPeopleSyntheticData2022}.

\begin{figure*}[t]
    \centering
        \subfigure[Not Visible]{
            \includegraphics[width=.3\linewidth]{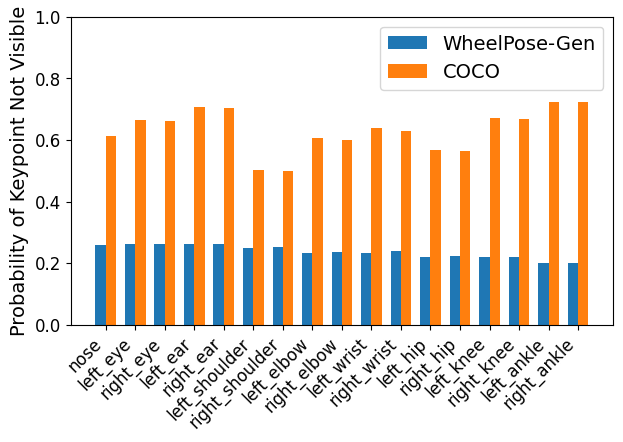}
            \label{subfig:NotVisibleKPProb}

            \Description{Bar graph showing the probability of keypoints being not visible on the Y axis and the 17 different COCO keypoints on the X axis. Two bars for each keypoint. COCO had a much probability for all keypoints, hovering around 0.5 while \systemname{}-Gen stayed around 0.2.}
        }
        \subfigure[Occluded]{
            \includegraphics[width=.3\linewidth]{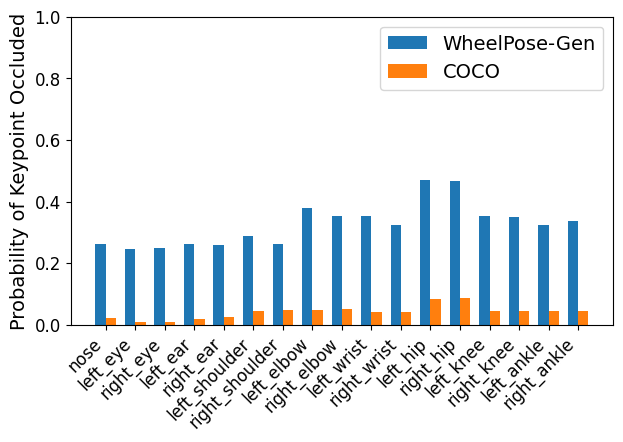}
            \label{subfig:OccludedKPProb}
            
            \Description{Bar graph showing the probability of keypoints being occluded on the Y axis and the 17 different COCO keypoints on the X axis. Two bars for each keypoint. \systemname{}-Gen had a much higher probability for all keypoints, hovering around 0.3 while COCO stayed around 0.05.}
        }
        \subfigure[Visible]{
            \includegraphics[width=.3\linewidth]{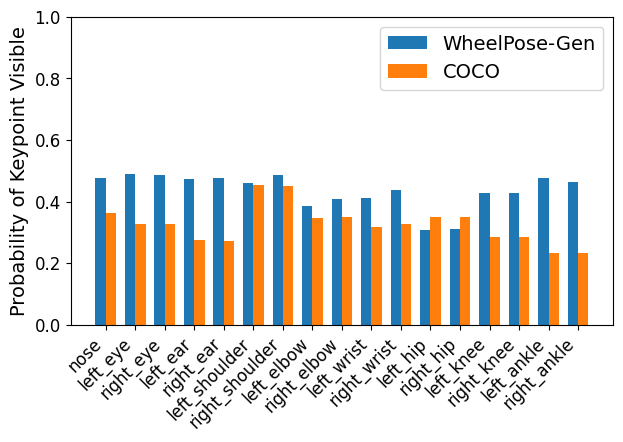}
            \label{subfig:VisibleKPProb}
            
            \Description{Bar graph showing the probability of keypoints being visible on the Y axis and the 17 different COCO keypoints on the X axis. Two bars for each keypoint. \systemname{}-Gen had a higher probability for most keypoints except the hips, hovering around 0.5 while COCO stayed around 0.4. \systemname{}-Gen hips had a probability of around .35.}
        }
    \vspace{-0.2cm}
    \caption{Probability of Occlusion Labelings per Keypoint. \Cref{subfig:NotVisibleKPProb} Probability of a keypoint being labeled as "not visible." 
    \Cref{subfig:OccludedKPProb} Probability of a keypoint being labeled as "occluded." 
    \Cref{subfig:VisibleKPProb} Probability of a keypoint being labeled as "visible." Labeling definitions are taken directly from the COCO occlusion and keypoint labeling standard. \cite{lin2015microsoft}.}
    \label{fig:KeypointOcclusionLabels}
\end{figure*}

\begin{figure*}
    \centering
    \includegraphics[width=.9\textwidth]{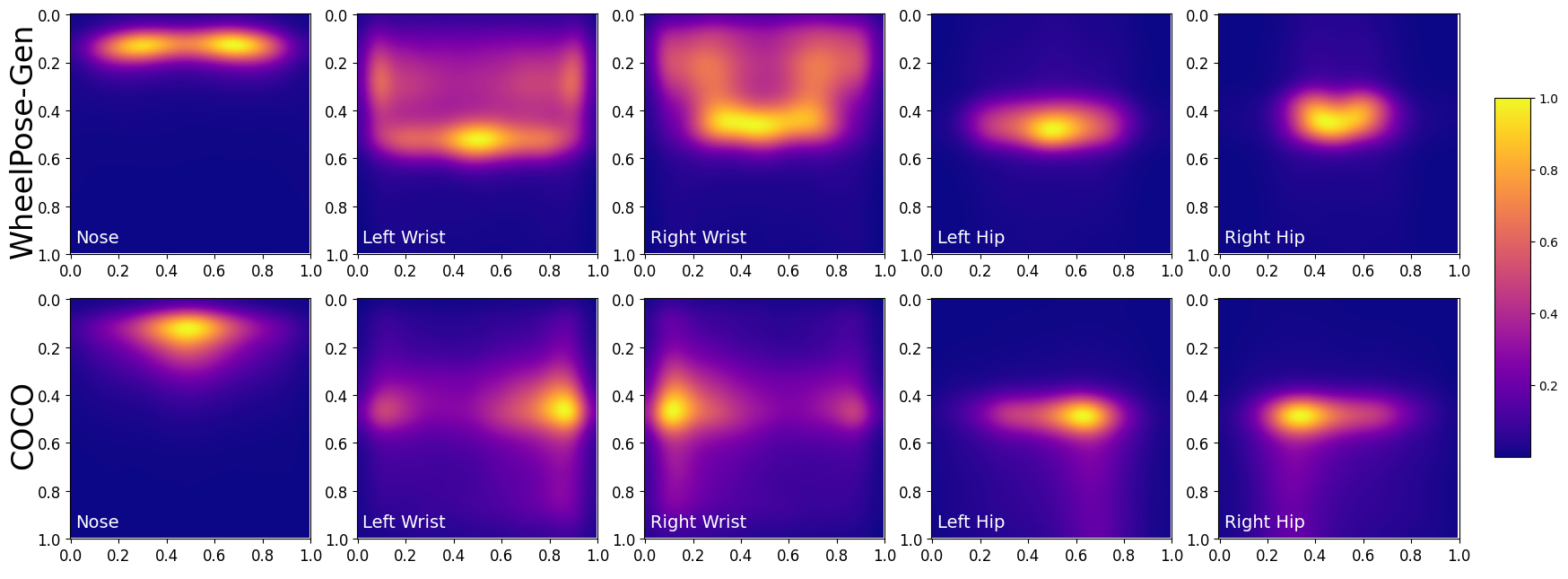}
    \vspace{-0.2cm}
    \caption{Heatmap of Five Keypoints Location. Top row: \systemname{}-Gen. Bottom row: COCO. All keypoints locations in the heatmap are computed by $(\frac{\text{x} - \text{bounding box top corner x}}{\text{bounding box width}}, \frac{\text{y} - \text{bounding box top corner y}}{\text{bounding box height}})$. The heatmap is normalized according to the size of the dataset.}
    \label{fig:KPLocHeatmap}

    \Description{Ten heatmaps depicting the spatial distribution of 5 keypoints with respect to the bounding box for \systemname{}-Gen and COCO. The distributions were as follows. Nose: \systemname{}-Gen and COCO both had a majority of the distribution fall at the top of the heatmap. Wrists: \systemname{}-Gen had a distribution across the entire top half of the heatmap while COCO centered around the left/right edge of the corresponding-sided wrist. COCO was also asymmetrical where the distribution peaked on the edge of the corresponding sided wrist. Hips: \systemname{}-Gen had hips centered in the middle of the heatmap with little deviation while COCO had hips centered around their corresponding side in the middle of the heatmap and deviated all the way to the bottom of the heatmap in a near T shape.}
\end{figure*}

Our dataset contains 296,508 human instances, of which 271,803 instances have annotated keypoint labeling. Note that not all human instances have annotated keypoints because some human instances had no joints within the camera's view despite still being visible. In comparison, COCO has 66,808 images with 273,469 human instances, of which 156,165 have annotated keypoints. This difference in human instances and annotated keypoints are both due to out-of-view keypoints like in our dataset, and a lack of keypoint labels due to human labeling errors. 

\subsubsection{Bounding Boxes}

We analyze the bounding box annotations by generating a set of statistics comparing \systemname{}-Gen against COCO (\Cref{fig:BBoxGeneral}) to evaluate the diversity of the frequency, placement, and size of human instances. We find that \systemname{}-Gen contains a more even distribution of the number of bounding boxes, or human instances, per image compared to COCO, indicating a greater diversity of images featuring wheelchair users in differently sized groups (\Cref{subfig:NBBoxes}). The COCO dataset does however have a higher concentration of images with large amounts of human instances within them, most likely due to the number of images depicting crowds of people. We also calculated the area of the bounding box relative to the overall image size to analyze the size of individual human instances (\Cref{subfig:BBoxArea}). Here, we observe that our bounding boxes have slightly more evenly distributed sizes in relation to the image size when compared to COCO. Note that our dataset consists of images of a uniform size ($1280\times720$), and COCO contains a wide variety of different image sizes up to a maximum size of $640\times640$. Thus, images with the same relative size in relation to the image dimensions are still a higher definition in \systemname{}-Gen comparatively.

We then analyze the spatial distribution of bounding boxes by plotting a heatmap of bounding box locations for both \systemname{}-Gen and COCO (\Cref{subfig:BBoxHeatmap}). We note that the location of bounding boxes is a direct product of the human generation, occluder, and camera randomizers and acts as a quantification of their effects. For both datasets, we overlay the bounding boxes with their location scaled by the overall size of the image. For the COCO dataset, we observe a majority of bounding boxes are centered in the middle of the image. We also observe that \systemname{}-Gen has a wider bounding box distribution with more spreading into the top edge of the image compared to COCO, indicating our randomization parameters create a more even spread of human instances across the image with more examples of humans at the edge of the camera. The center of the distribution of bounding boxes in \systemname{}-Gen is also slightly higher in the image than that of COCO. 

\subsubsection{Keypoints}

We first measure the probability of a keypoint to be one of the three predefined COCO occlusion levels (not visible, occluded, visible) in \systemname{}-Gen and COCO as further quantification of the effects of the randomizers listed previously. In the context of \systemname{}, not visible is when a keypoint is not in the image and has no prediction, occluded is when a keypoint is in the image but not visible (e.g., behind an object), and visible is when a keypoint is seen in an image.  Here we see that \systemname{}-Gen displays a significantly smaller probability of having nonvisible keypoints and a more uniform distribution of keypoints compared to COCO (\Cref{subfig:NotVisibleKPProb}). We also notice that \systemname{}-Gen has a far higher probability for a keypoint to be labeled as occluded compared to COCO, especially in the hips (\Cref{subfig:OccludedKPProb}). This can be explained by the different methods both datasets used in keypoint annotation. While \systemname{}-Gen uses a self-occlusion labeler defined in PSP which computes the distance between each keypoint and the closest visible part of the object within a threshold to determine occlusion labeling \cite{ebadiPeopleSansPeopleSyntheticData2022}, COCO is a fully human-labeled dataset. Within human annotators, there are variations between how an annotator might define and classify as occluded and not visible. \systemname{} does not suffer from the same issue as the labeler has information on the full 3D scene and can precisely identify every keypoint location over a human annotator who only has access to a single 2D view with no additional context. This phenomenon can be further seen in the probability of keypoint visibility, where \systemname{}-Gen displayed a higher probability for all keypoints except the two keypoints on hips (\Cref{subfig:VisibleKPProb}) where many hip keypoints are labeled as occluded due to the self-occlusion of the wheelchair.

We evaluate the diversity of our poses by creating a heatmap of keypoint annotations locations scaled by the corresponding bounding box in \systemname{}-Gen and COCO (\Cref{fig:KPLocHeatmap}). \systemname{}-Gen displays a wider distribution of potential keypoint locations in upper body keypoints compared to COCO. Lower body keypoints display a smaller distribution due to the limitations of postures in a wheelchair. We see for asymmetrical keypoints (left/right wrist), \systemname{}-Gen is far more evenly distributed across the X axis compared to COCO, a clear indication of a more even distribution of front, side, and back-facing human instances. We also notice a smaller Y-axis distribution in \systemname{}-Gen due to the presence of the wheelchair limiting potential movements up and down.

\subsubsection{Camera}

\begin{figure}
    \centering
        \subfigure[Camera Elevation]{
            \includegraphics[width=.45\linewidth]{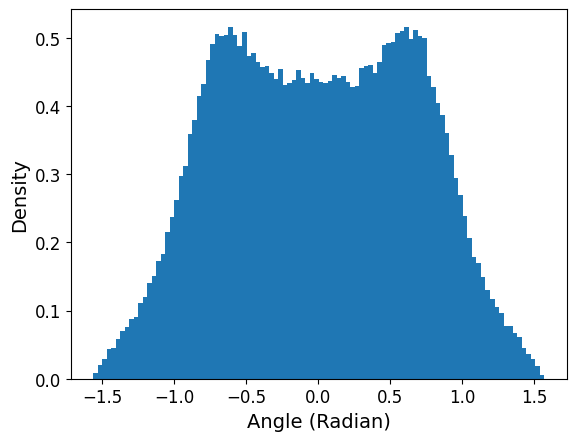}
            \label{subfig:CamElevation}

            \Description{Histogram of camera elevation with density from 0 to 0.5 on the Y axis and angles in radians from -1.57 to 1.57 on the X axis. The distribution roughly follows a normal distribution but peaks twice to 0.5 density at -0.5 and 0.5 radians before dipping to 0.4 at 0 radians.}
        }
        \subfigure[Camera Azimuth]{
            \includegraphics[width=.45\linewidth]{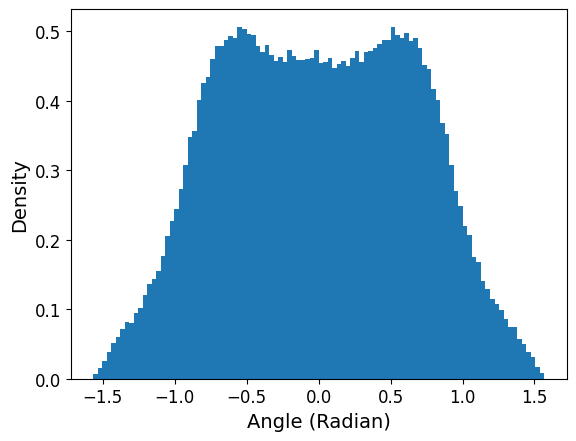}
            \label{subfig:CamAzimuth}     
            
            \Description{Histogram of camera azimuth with density from 0 to 0.5 on the Y axis and angles in radians from -1.57 to 1.57 on the X axis. The distribution roughly follows a normal distribution but peaks twice to 0.5 density at -0.5 and 0.5 radians before dipping to 0.45 at 0 radians.}

        }
        \subfigure[Camera Distance]{
            \includegraphics[width=.45\linewidth]{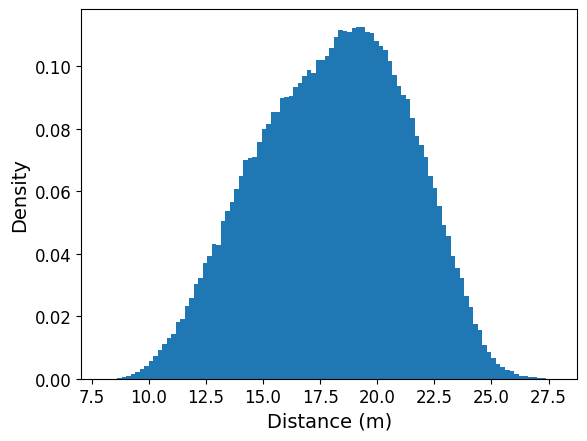}
            \label{subfig:CamDistance} 
            
            \Description{Histogram of camera distance with density from 0 to 0.12 on the Y axis and distance in meters from 7.5 to 27.5 on the X axis. The distribution roughly follows a normal distribution peaking at 20 meters with a density of about 0.11.}

        }
        \hspace{.5cm}
        \subfigure[Camera Potential Angles]{
            \includegraphics[width=.38\linewidth]{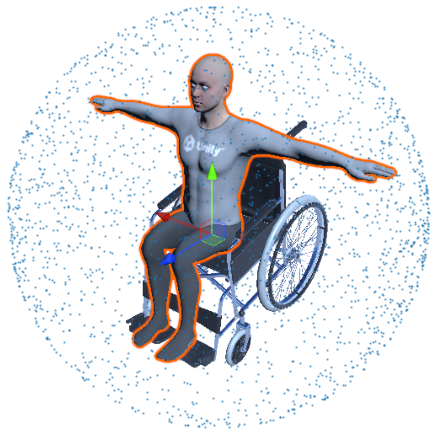}
            \label{subfig:CamAngles}

            \Description{Visualization of potential camera angles. A human model in a wheelchair in T-pose is surrounded by a set of points on the surface of a sphere representing a potential camera angle. The human model has camera angles surrounding all sides of it, creating an evenly distributed sphere of camera angles.}
        }
        \caption{Distribution of Potential Camera Angles and Distances. 
        \Cref{subfig:CamElevation} Distribution of elevation angle (up-down, positive indicating a camera above the nose and looking down). 
        \Cref{subfig:CamAzimuth} Distribution of azimuth angle (left-right, positive indicating a camera is to the right looking left) distribution. 
        \Cref{subfig:CamDistance} Distribution of camera distance to a human instance. 
        \Cref{subfig:CamAngles} Visualization of potential camera angles. Computed by sampling camera location relative to human instance every 100 human instances and visualizing the corresponding unit vector.}
        \label{fig:CamEvaluations}
\end{figure}

Finally, we quantify the variations in our camera placement and rotation. Recent studies have shown the critical impact of diverse camera angles on model performance in 3D human recovery problems \cite{cai2022playing, madan2021cnns, park2023MultiView}. We visualize our diversity of camera angles and distances and observe a wide distribution of potential elevations, azimuths, and distances (\Cref{fig:CamEvaluations}). We then sample a set of camera locations relative to individual human instances and visualize their angle around the instance to observe full coverage of camera angles all around instances (\Cref{subfig:CamAngles}). All visualizations indicate a loosely followed Gaussian distribution with a wide variety of different camera angles. We do not compare these statistics with COCO since it does not include camera configuration information. Our camera configuration parameters can be found in \Cref{tab:EnvironmentParameters}. 

\subsection{Model Performance Evaluation}
\label{sec:ModelPerformance}

\subsubsection{Testing Dataset}

There currently does not exist an image dataset focused primarily on wheelchair users. For the testing of our system, we create a new dataset of 2,464 images of wheelchair users collected from 84 \textit{YouTube} videos in a similar data collection process as other computer vision works \cite{andriluka14cvpr, wang2019Youtube}. A set of 16 action classes consisting of common daily tasks wheelchair and able-bodied users both perform and unique wheelchair sports were selected (e.g., talking, basketball, rugby, dancing, etc.). Action classes are listed in \Cref{tab:ActivityClasses}. Videos are collected through keyword searches revolving around each of the action classes. Annotators iterate through 500 equally spaced frames (minimum one-second intervals) from each video and identify frames with poses and settings sufficiently different from the previously collected images. Annotators ensure that there is a wheelchair user within view for every collected frame. Crowd worker involvement is then utilized to annotate bounding boxes and keypoint locations on collected images. Researchers manually validated results from this process for accuracy. Examples of this dataset can be found in our open-source repository at \url{https://github.com/hilab-open-source/wheelpose}.

\begin{table*}[]
\caption{Ablation trials with configurations of each dataset and results in terms of the mean Average Precision (mAP) in the bounding box (BB) and keypoint (KP) detection tasks.}
\label{tab:Ablation}
\begin{tabular}{@{}llllll@{}}
\toprule
    \textbf{Dataset} & 
    \textbf{Human Model} & 
    \textbf{Background} & 
    \textbf{Animation} & 
    \textbf{BB mAP} & 
    \textbf{KP mAP} \\ \midrule
        \changed{\systemname{}-base} & 
        \changed{PSP default} & 
        \changed{PSP default} & 
        \changed{HumanML3D} & 
        \changed{68.68} & 
        \changed{65.18} \\
        \changed{\systemname{}-SH} & 
        \changed{SyntheticHumans} &
        \changed{PSP default} & 
        \changed{HumanML3D} & 
        \changed{68.53} & 
        \changed{\textbf{68.04}} \\ 
        \systemname{}-t2m & 
        PSP default & 
        PSP default & 
        Text2Motion & 
        68.95 & 
        64.83 \\
        \systemname{}-t2mR10 & 
        PSP default & 
        PSP default & 
        Text2Motion random 10\% removed & 
        68.97 & 
        65.19 \\
        \systemname{}-t2mHE10 & 
        PSP default & 
        PSP default & 
        Text2Motion HE 10\% removed    & 
        \changed{69.36} & 
        \changed{65.53} \\
        \changed{\systemname{}-SB} & 
        \changed{PSP default} & 
        \changed{SynthHomes} & 
        \changed{HumanML3D} & 
        \changed{\textbf{69.75}} & 
        \changed{66.60} \\
        \changed{\systemname{}-RB} & 
        \changed{PSP default} & 
        \changed{BG-20K} & 
        \changed{HumanML3D} & 
        \changed{64.44} & 
        \changed{63.89} \\ \midrule
        \systemname{}-Gen & 
        SyntheticHumans & 
        SynthHomes & 
        Text2Motion & 
        \textbf{69.71} & 
        67.53 \\ 
        \changed{\systemname{}-Opt} &
        \changed{SyntheticHumans} &
        \changed{SynthHomes} &
        \changed{Text2Motion HE 10\% removed} &
        \changed{69.58} &
        \changed{\textbf{67.96}} \\ \midrule
        ImageNet (baseline) & 
        N.A. & 
        N.A. & 
        N.A. & 
        \textbf{35.19} & 
        \textbf{63.11} \\
        PSP (baseline) & 
        PSP Default & 
        PSP Default & 
        PSP Default Able-Bodied & 
        31.03 & 
        53.26 \\ \bottomrule
\end{tabular}
\end{table*}

\subsubsection{Training Strategy}
\label{subsubsec:TrainingStrategy}
Similar to training outlined in PSP-HDRI \cite{ebadiPSPHDRISyntheticDataset2022}, all of our models in this evaluation utilized ResNet-50 \cite{7780459} plus Feature Pyramid Network (FPN) \cite{lin2017feature} backbones. Additionally, these models were fine-tuned using the starting weights and framework of the Detectron2 ImageNet Keypoint R-CNN R50-FPN variant \cite{he2018mask}.

We create each model in the same way: fine-tuning with the relevant dataset on the backbone described previously. We opted for a simple training approach, setting the initial learning rate to 0.000025 for 20 epochs, and then lowering the learning rate by a factor of 10 for an additional 10 epochs. For the first 1000 iterations we also conduct a linear warm-up of the learning rate to its starting value, slowly increasing the learning rate to its starting value. The momentum was set to 0.9 and the weight decay was set to 0.0001. All training runs were completed using a 4.2GHz 16-core/32-thread AMD Ryzen Threadripper PRO 3955WX CPU, 2 NVIDIA RTX A5500 24GB VRAM, and 256 GB 3200MHz DDR4 memory with a mini-batch size of 13 images per GPU, where each image was normalized using the mean pixel value and standard deviation of the ImageNet base model. For each model, we checkpointed the model weights during every epoch and selected the epoch with the best-performing keypoint AP to report in evaluation.
This evaluation scheme was consistent across baseline datasets and our synthetic datasets.  

We note that \systemname{} fine-tuned models and individual human evaluators may not agree with each other on where a keypoint is due to the innate difference between a modeled person and a real person. Changes in the way \systemname{} defines keypoints can alter where a keypoint is predicted and the metrics computed in the coming sections as seen in \Cref{sec:KeypointLocation}. We attempted to minimize these differences as much as possible through realistic keypoint definitions in Unity. 

\subsubsection{Ablation Testing Strategy}

We address \textbf{(RQ2.1)} through ablation testing\footnote{Ablation testing involves the removal of certain components to understand the contribution of the component to the overall performance of an AI system} on a set of selected domain randomization parameters, including animations, backgrounds, and human models, to better understand the performance impacts of select data generation parameters.

\noindpar{Configuration.}
Regarding the \textit{human model}, we analyze the effects on the model performance of using: 1) PSP default human models (PSP Default), and 2) SyntheticHuman human models (SyntheticHumans).
As to the \textit{background}, we compare between 1) PSP default texture backgrounds (PSP Default), 2) SynthHomes background images (SynthHomes), and 3) BG-20K real background images (BG-20K).
Finally, regarding the \textit{motion sequence}, we compare between 1) HumanML3D animations (HumanML3D), 2) Text2Motion animations (Text2Motion), 3) Text2Motion with $10\%$ of animations randomly removed (Text2Motion random $10\%$ removed), and 4) Text2Motion animations with the bottom $10\%$ of animations in total ease and frequency from human evaluations (HE) (Text2Motion HE $10\%$ removed). \Cref{tab:Ablation} shows this list of datasets and their configurations.

We use combinations of these parameters to assemble a set of datasets of 70,000 images each. Each dataset consisted of $\approx65,000$ images with at least one human instance and was used to fine-tune the base ImageNet model using the strategy listed in \Cref{subsubsec:TrainingStrategy}. We also include the original PSP dataset with $\approx65,000$ images with at least one human instance generated from the provided Unity environment to test the efficacy of fine-tuning with synthetic able-bodied user data \cite{ebadiPeopleSansPeopleSyntheticData2022}. For these tests, all models and our baseline tests, ImageNet and PSP, were tested on our real wheelchair data testing set using the industry standard metric for detection and estimation accuracy, COCO mean Average Precision \cite{lin2015microsoft}.

\noindpar{Results.}
Results in terms of bounding box (BB) and keypoint (KP) mean APs (mAP) across ablation trials are shown in \Cref{tab:Ablation}. 

Regarding person detection (BB) performance, all datasets demonstrated significant performance boosts of varying degrees when compared to ImageNet and PSP. Notably, our best-performing ablation test led to a  $98.21\%$ improvement in BB mAP (\systemname{}-SB) and a $7.81\%$ improvement in keypoint mAP (\systemname{}-SH) over the best baseline dataset (ImageNet). This indicates the efficacy of our synthetic data and the large headroom for improvement in detecting poses of wheelchair users with industry-standard deep learning models.

\begin{table*}[]
\caption{Bounding box AP performance comparison between base models and \systemname{}-Opt. We list the mean of our seeded testing $\pm$ the maximum absolute deviation from the mean.}
\label{tab:BBAP}
\begin{tabular}{@{}lllllll@{}}
\toprule
  \textbf{Dataset} &
  \textbf{BB mAP} &
  \textbf{BB AP\textsuperscript{\textit{IoU=.50}}} &
  \textbf{BB AP\textsuperscript{\textit{IoU=.75}}} &
  \textbf{BB AP\textsuperscript{\textit{small}}} &
  \textbf{BB AP\textsuperscript{\textit{medium}}} &
  \textbf{BB AP\textsuperscript{\textit{large}}} \\ \midrule
    \changed{\systemname{}-Opt}   & 
    \changed{$69.46\pm0.22$} & 
    \changed{$90.77\pm0.20$} & 
    \changed{$82.75\pm0.61$} & 
    \changed{$3.24\pm0.76$} & 
    \changed{$59.82\pm0.85$} & 
    \changed{$69.91\pm0.21$} \\ \midrule
    ImageNet (baseline) & 
    35.19 & 
    71.49 & 
    28.50 & 
    0.00 & 
    3.15 & 
    36.12 \\
    PSP (baseline) & 
    31.03 & 
    63.64 & 
    25.08 & 
    0.00 & 
    6.27 & 
    32.35 \\ \bottomrule
\end{tabular}
\end{table*}

\begin{table*}[]
\caption{Keypoint AP performance comparison between baseline and \systemname{}-Opt. We list the mean of our seeded testing $\pm$ the maximum absolute deviation from the mean. We do not include AP\textsuperscript{\textit{small}} since the human is too small to accurately assign keypoints.}
\label{tab:KPAP}
\begin{tabular}{@{}llllll@{}}
\toprule
    \textbf{Dataset} & 
    \textbf{KP mAP} & 
    \textbf{KP AP\textsuperscript{\textit{OKS=.50}}} & 
    \textbf{KP AP\textsuperscript{\textit{OKS=.75}}} & 
    \textbf{KP AP\textsuperscript{\textit{medium}}} & 
    \textbf{KP AP\textsuperscript{\textit{large}}} \\ \midrule
        \changed{\systemname{}-Opt}   & 
        \changed{$67.93\pm0.02$} & 
        \changed{$87.61\pm0.19$} & 
        \changed{$74.48\pm0.25$} & 
        \changed{$35.48\pm0.40$} & 
        \changed{$68.99\pm0.06$} \\ \midrule
        ImageNet (baseline) & 
        63.11 & 
        77.43 &
        67.20 & 
        6.22  & 
        64.96 \\
        PSP (baseline) & 
        53.26 & 
        68.07 & 
        57.36 & 
        10.15 & 
        56.12 \\ \bottomrule
\end{tabular}
\end{table*}

Poor performance from BG-20K may be explained by the detail of background images. PSP default and SynthHomes data tend to feature a set of simple or smooth textures while real-world images are often more detailed and consist of more texture. These results may signal a preference for other background characteristics over pure realism.

When examining keypoint performance, the most significant improvement was the inclusion of Unity SyntheticHumans models. This makes intuitive sense, as the diverse and more representative human models more closely match up with the humans found in the real world rather than the generalized Unity humanoid models. The variations between models also help combat overfitting issues by introducing many different definitions of what is a "human" to the model.

Finally, we examine the performance tradeoffs in our different animation sets. We found motion generation outputs with random filtering  (\systemname{}-t2mR10) performed comparably to motion capture data (\systemname{}), indicating similar motion quality between the two data sources. We also found that randomly filtered Text2Motion animations (\systemname{}-t2mR10) performed better than the full motion set (\systemname{}-t2m). This may imply that the number of animations and poses is not directly correlated with model performance. It is important to note these results may change depending on what animations have been filtered. This idea is further shown in the HE model (\systemname{}-t2mHE10), which showed improvement in both BB and KP mAP over the randomly removed $10\%$ model. We see that the removal of specific animations that are not perceived as "realistic" can improve the model performance of generated data.

\changed{From our ablation testing results, we assembled two new datasets: \systemname{}-Gen, a dataset created from fully generative parameters using SyntheticHuman models, SynthHomes backgrounds, and all $100$ Text2Motion animations, and \systemname{}-Opt, a dataset created from the best performing parameters from ablation testing which include SyntheticHuman models, SynthHomes backgrounds, and Text2Motion HE 10\% removed animations. Both datasets performed comparably to the best performing ablation test in BB and KP mAP (\systemname{}-SH) with \systemname{}-Opt. We note that \systemname{}-Opt outperforms \systemname{}-Gen in KP mAP which follows our findings in the initial ablation test. Our findings indicate that different combinations of domain randomization parameters can produce better AI models than the best perform parameters individually.}

\subsubsection{\changed{\systemname{}-Opt Model Performance Deep Dive}}

We further evaluate \textbf{(RQ2.2)} by conducting an in-depth quantitative and qualitative analysis of the changes in performance in Detectron2 when fine-tuned with \systemname{}-Opt, the best performing dataset from ablation testing using synthetic data and simple human evaluations. 

\noindpar{Configuration.}
We trained and evaluated the results for \systemname{}-Opt with the same strategy described in \Cref{subsubsec:TrainingStrategy} three separate times using different model seeds (42, 4242, 424242). We then computed the mean and maximum absolute deviation of a set of evaluation metrics, including BB AP, KP AP, and individual keypoint metrics, across the three trials. We compute the same metrics on ImageNet and PSP for use as our baseline.

\noindpar{Results.}
We first quantify our overall bounding box and keypoint performance with AP and its related breakdowns to build an overarching view of our dataset's performance across different scenarios. AP at different IoU and OKS\footnote{IoU and OKS perform the same fundamental purpose for bounding boxes and keypoints respectively} thresholds measure the prediction accuracy at varying degrees of recall (Higher indicates a stricter ground truth definition). Additionally, the overall AP score is split into small, medium, and large based on the detection segment area to quantify the performance at different camera distances and human instance sizes. More information is found in the COCO documentation \cite{lin2015microsoft}.

\Cref{tab:BBAP} lists the BB AP performance for \systemname{}-Opt and the baselines. Our dataset displays over a 98\% improvement over both baseline scores in all subcategories of AP. Furthermore, we notice a major drop in performance in the baseline models as the IoU threshold becomes stricter. In contrast, our \systemname{} models maintain a high level of accuracy across a much wider range of IoU thresholds. These results indicate \systemname{}-Opt can not only identify a wheelchair user but is capable of drawing an accurate bounding box around the entire human instance compared to the baseline models which are only capable of low-fidelity wheelchair user detections.

As shown in \Cref{tab:KPAP}, pose estimation improved by up to 7.64\% in KP mAP in \systemname{}-Opt when compared to ImageNet and up to 27.54\% when compared to PSP. Furthermore, we see a similar or greater magnitude of improvement across all AP metrics, indicating the use of \systemname{}-Opt improves pose estimation in all scenarios. We note drastic improvements in AP on medium-sized human instances, where ImageNet had noticeably poor performance (6.22). Our system, thus, not only improves but even enables existing models to estimate the postures of wheelchair chairs at further distances.

\begin{table*}[]
\caption{Keypoint performance breakdown of our primary datasets. We list the mean of our seeded testing $\pm$ the maximum absolute deviation from the mean. Within the parentheses on the right is the percent difference from the base model, ImageNet. Percentage of Detected Joints (PDJ) @ 5 describes the percentage of correctly predicted joints within a 5\% bounding box diagonal radius \cite{sapp2013modec}. Per Joint Position Error (PJPE) describes the mean Euclidean distance for each keypoint from the ground truth \cite{zwolfer2023Ski}. Object Keypoint Similarity Score (OKS) as described by COCO \cite{lin2015microsoft} and used in Chen \etal \cite{chen2021transfer} is the mean precision per keypoint evaluated at both standard loose and strict similarity thresholds of 0.5 and 0.75 respectively. Superiority directions are noted as + and - next to each metric. }

\label{tab:KPPerformanceBreakdown}
\begin{tabular}{@{}lllll@{}}
\toprule
  \textbf{Keypoint} &
  \textbf{PDJ@5 (+)} &
  \textbf{PJPE (-)} &
  \textbf{OKS50 (+)} &
  \textbf{OKS75 (+)} \\ \midrule
	nose &
	$0.91\pm0.02 (-2.15\%)$ &
	\enspace$8.73\pm4.06 (-77.38\%)$ &
	$0.90\pm0.02 (-3.23\%)$ &
	$0.89\pm0.03 (-1.83\%)$ \\
	eyes &
	$0.99\pm0.00 (+5.70\%)$ &
	\enspace$6.30\pm0.64 (-77.6\%)$ &
	$0.96\pm0.01 (+3.76\%)$ &
	$0.95\pm0.01 (+4.19\%)$ \\
	ears &
	$0.92\pm0.01 (+5.56\%)$ &
	$10.90\pm0.03 (-61.60\%)$ &
	$0.93\pm0.00 (+5.90\%)$ &
	$0.83\pm0.01 (+0.81\%)$ \\
	shoulders &
	$0.92\pm0.02 (+6.74\%)$ &
	$10.98\pm0.26 (-63.89\%)$ &
	$1.00\pm0.01 (+4.74\%)$ &
	$0.97\pm0.01 (+6.38\%)$ \\
	elbows &
	$0.86\pm0.01 (+2.78\%)$ &
	$14.54\pm0.23 (-69.38\%)$ &
	$0.97\pm0.00 (+5.80\%)$ &
	$0.92\pm0.00 (+6.55\%)$ \\
	wrists &
	$0.85\pm0.01 (-0.19\%)$ &
	$16.22\pm1.65 (-76.22\%)$ &
	$0.92\pm0.02 (+6.13\%)$ &
	$0.88\pm0.02 (+1.73\%)$ \\
	hips &
	$0.65\pm0.01 (-18.96\%)$ &
	$22.96\pm0.61 (-51.24\%)$ &
	$0.97\pm0.00 (-2.02\%)$ &
	$0.89\pm0.01 (-6.63\%)$ \\
	knees &
	$0.89\pm0.02 (+6.63\%)$ &
	$14.99\pm0.97 (-78.85\%)$ &
	$0.97\pm0.01 (+5.25\%)$ &
	$0.94\pm0.01 (+7.01\%)$ \\
	ankles &
	$0.78\pm0.01 (+76.14\%)$ &
	$18.16\pm0.95 (-69.04\%)$ &
	$0.94\pm0.00 (+87.67\%)$ &
	$0.90\pm0.01 (+91.49\%)$ \\ \bottomrule
\end{tabular}
\end{table*}

Finally, we compute a set of per-keypoint metrics shown in \Cref{tab:KPPerformanceBreakdown} to analyze the differences in specific keypoint predictions between \systemname{}-Opt and the baseline. We only showcase percent change with respect to ImageNet as the PSP fine-tuned model performs drastically worse in nearly all BB and KP metrics. Upon examining the percentage of detected joint (PDJ) \cite{sapp2013modec} at a 5\% threshold, a measure of a model's ability to identify a joint, we notice a 76.14\% improvement in ankle detection, attributed to more information on foot placement in wheelchairs, and a 18.96\% decrease in detected hips, attributed to the wheelchair obstructing most of the lower torso and hip area. We then compute the per joint position error (PJPE), a simple accuracy metric measuring the Euclidean distance error in detected joints, and found that as long as a joint is detected, \systemname{}-Opt predicts keypoints over 51.24\% (hips) more accurately. Finally, we compute the meaned per-keypoint precision values at OKS thresholds of 0.5 and 0.75 as another measure of individual keypoint prediction accuracy. We see slight improvements across most joints. Similar to PDJ, we can see similar trends in the ankles and hips, improving a significant amount or slightly worse respectively. 

\subsubsection{Key Prediction Changes}
We conduct a visual analysis of the changes in predicted keypoints between ImageNet and \systemname{}-Opt to analyze the information transfer between our synthetic data and the real world across different scenarios to identify specific situations where we perform better or worse. We ignore the PSP dataset as it demonstrated noticeably worse performance compared to ImageNet in both bounding box and pose estimation (\Cref{tab:BBAP,tab:KPAP}). Examples of different trends were plotted with the predictions from both \systemname{}-Opt, in green, and ImageNet, in red, overlaid on top.

\begin{figure*}
    \subfigure[]{
        \includegraphics[width=.18\linewidth]{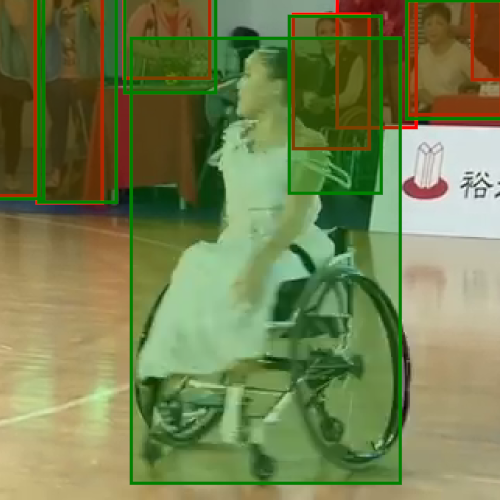}
        \label{subfig:DetectionImprovements1}

        \Description{A wheelchair dancer gets ready for her performance in front of a crowd at a 45-degree angle from the camera.}
    }
    \subfigure[]{
        \includegraphics[width=.18\linewidth]{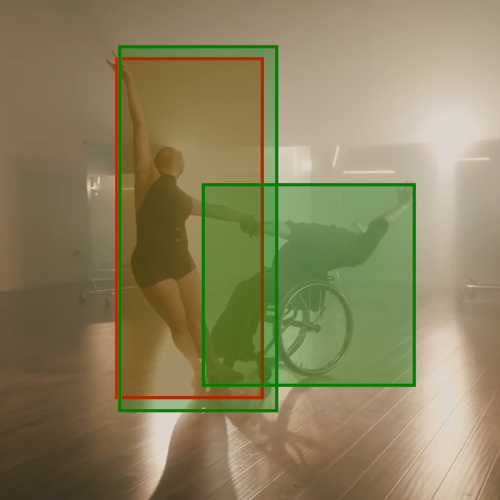}
        \label{subfig:DetectionImprovements2}

        \Description{A wheelchair leans back while holding hands with an able bodied dancer.}
    }
    \subfigure[]{
        \includegraphics[width=.18\linewidth]{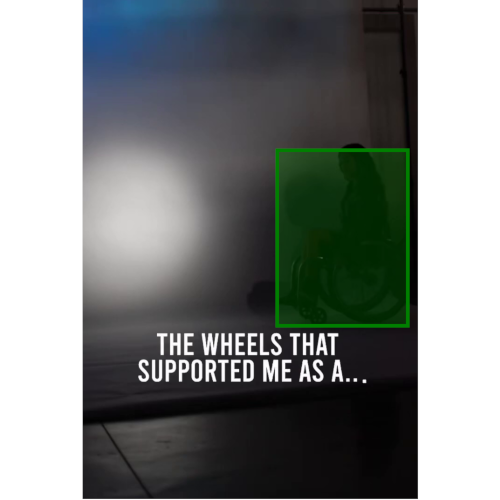}
        \label{subfig:DetectionImprovements3}

        \Description{A side view of a wheelchair user in a dark lighting condition against a projector.}
    }
    \subfigure[]{
        \includegraphics[width=.18\linewidth]{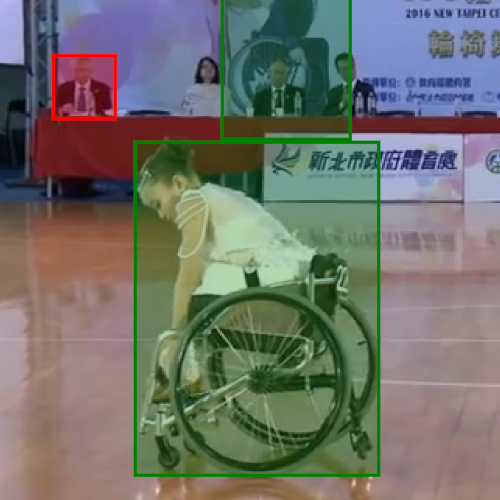}
        \label{subfig:DetectionImprovements4}

        \Description{A side view of a wheelchair dancer with their back hunched down. In the background, there are four judges against a promotional poster featuring a wheelchair user.}
    }
    \subfigure[]{
        \includegraphics[width=.18\linewidth]{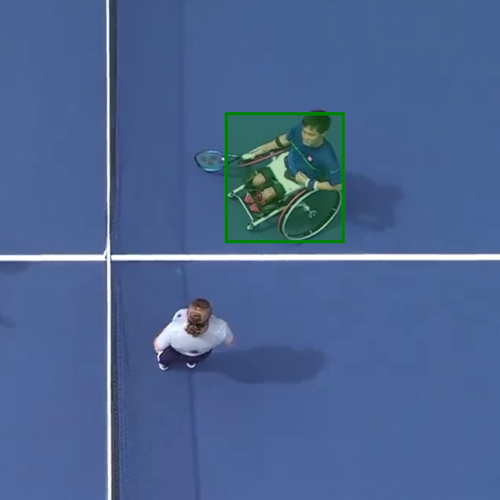}
        \label{subfig:DetectionImprovements5}

        \Description{A top down view of a wheelchair tennis player in a court. The wheelchair user is holding a tennis racket. Another person is standing on the court.}
    }
    \subfigure[]{
        \includegraphics[width=.18\linewidth]{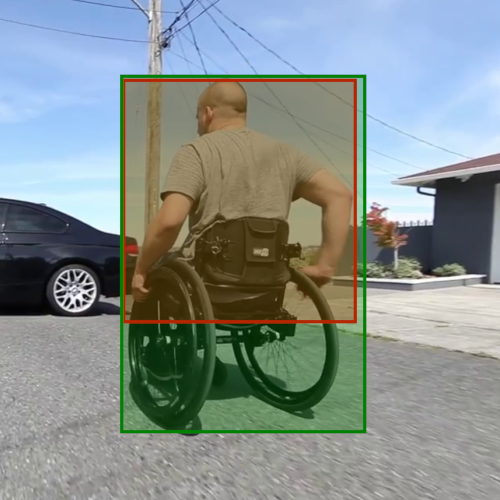}
        \label{subfig:DetectionImprovements6}

        \Description{A back view of a wheelchair user going towards a car. Both hands are on their wheels as they push towards the vehicle.}
    }
    \subfigure[]{
        \includegraphics[width=.18\linewidth]{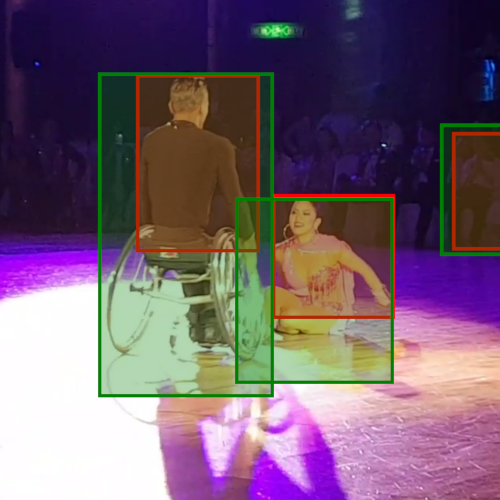}
        \label{subfig:DetectionImprovements7}

        \Description{A back view of a wheelchair user dancing with an able-bodied user. The able bodied user is sitting on the ground as the wheelchair user looks at them.}
    }
    \subfigure[]{
        \includegraphics[width=.18\linewidth]{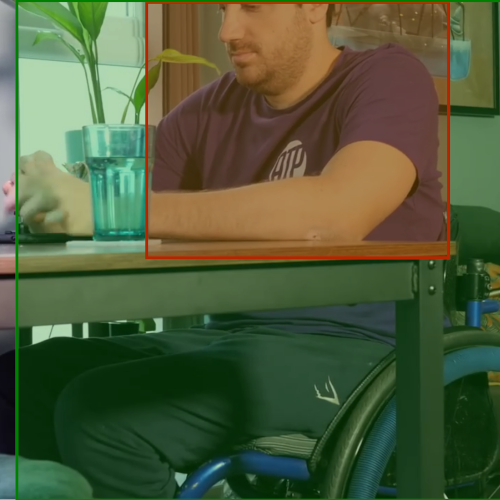}
        \label{subfig:DetectionImprovements8}

        \Description{A close-up 45-degree front view of a wheelchair user at a table. The camera cuts off part of their head as they use their phone. The camera captures their legs under the table.}
    }
    \subfigure[]{
        \includegraphics[width=.18\linewidth]{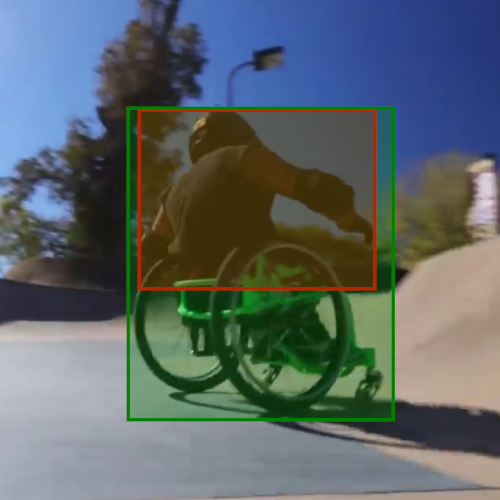}
        \label{subfig:DetectionImprovements9}

        \Description{A back-view of a wheelchair user entering a slope at a skate park. The user is pushing fast with both hands in front of them. They are wearing a helmet and elbow pads.}
    }
    \subfigure[]{
        \includegraphics[width=.18\linewidth]{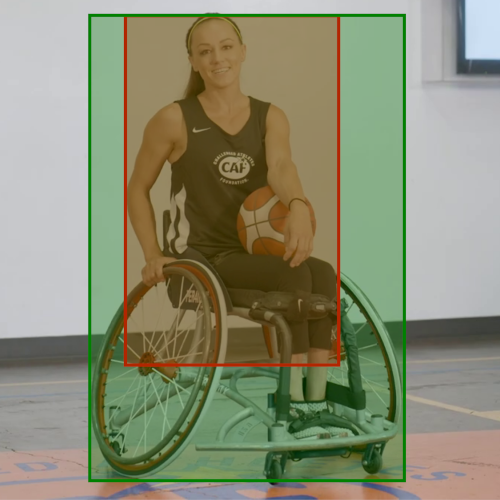}
        \label{subfig:DetectionImprovements10}

        \Description{A front view of a wheelchair user holding a basketball in their lap. They are looking directly at the camera.}
    }
    \vspace{-0.5cm}
    \caption{Examples of wheelchair user detection improvements with \systemname{}-Opt over ImageNet. Red represents ImageNet predictions while green represents \systemname{}-Opt predictions. \Cref{subfig:DetectionImprovements1,subfig:DetectionImprovements2,subfig:DetectionImprovements3,subfig:DetectionImprovements4,subfig:DetectionImprovements5} all show wheelchair users in different scenarios who were completely undetected by ImageNet but detected with \systemname{}-Opt fine-tuning. \Cref{subfig:DetectionImprovements6,subfig:DetectionImprovements7,subfig:DetectionImprovements8,subfig:DetectionImprovements9,subfig:DetectionImprovements10} all show wheelchair users in different scenarios who were detected by ImageNet, but had poor bounding box predictions which were improved in \systemname{}-Opt fine-tuning.}
    \vspace{-0.3cm}
    \label{fig:DetectionImprovements}
\end{figure*}

\noindpar{Improvements in wheelchair user detection.}
As shown in the BB AP improvements between ImageNet and \systemname{}-Opt in \Cref{tab:BBAP}, we notice major improvements in wheelchair user detection. \Cref{fig:DetectionImprovements} shows some examples of these improvements in a variety of different environments. \Cref{subfig:DetectionImprovements1,subfig:DetectionImprovements4,subfig:DetectionImprovements5} shows two examples of proper wheelchair user detection through \systemname{}-Opt. \Cref{subfig:DetectionImprovements2,subfig:DetectionImprovements3} shows two examples of wheelchair users being detected in low visibility settings due to both extremely bright and dark lighting conditions. Of particular note is \Cref{subfig:DetectionImprovements4}, where even the background poster of a wheelchair image has been detected, which human annotators had missed. We also notice that even if ImageNet had detected a wheelchair user, its bounding box prediction still was not accurate. \Cref{subfig:DetectionImprovements6,subfig:DetectionImprovements7,subfig:DetectionImprovements8,subfig:DetectionImprovements9,subfig:DetectionImprovements10} shows some examples of this, where ImageNet tends to cut off portions of the wheelchair user's full body while \systemname{}-Opt captures a more accurate and representative bounding box.

\begin{figure*}
    \subfigure[]{
        \includegraphics[width=.23\linewidth]{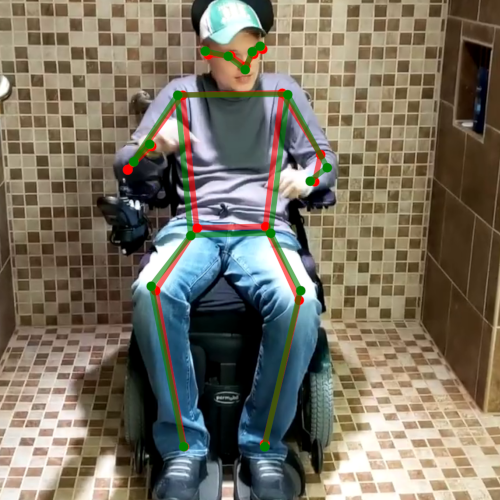}
        \label{subfig:SimilarPerformance1}

        \Description{A wheelchair user in a motorized wheelchair facing forward in the middle of a tiled room. The user's arms are moving in front of them.}
    }
    \subfigure[]{
        \includegraphics[width=.23\linewidth]{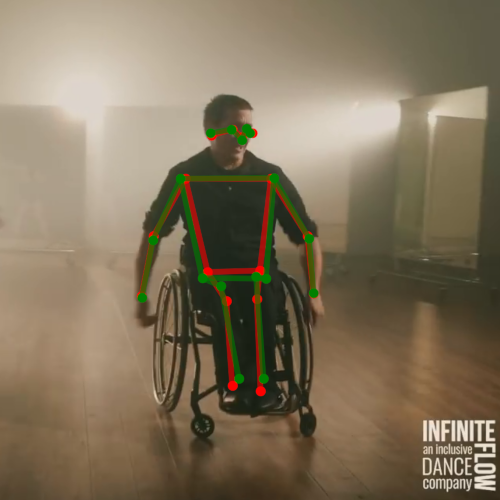}
        \label{subfig:SimilarPerformance2}

        \Description{A wheelchair user pushing forward towards the camera in the middle of a bright room with mirrors. The overall brightness of the image is high.}
    }
    \subfigure[]{
        \includegraphics[width=.23\linewidth]{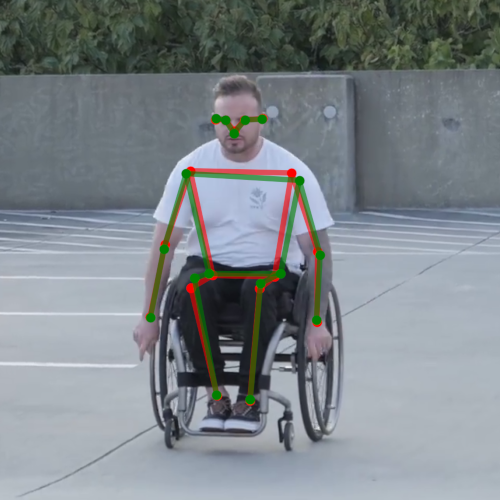}
        \label{subfig:SimilarPerformance3}

        \Description{A wheelchair user pushing forward towards the camera in a parking lot. Both hands are placed on the front of the wheel.}
    }
    \subfigure[]{
        \includegraphics[width=.23\linewidth]{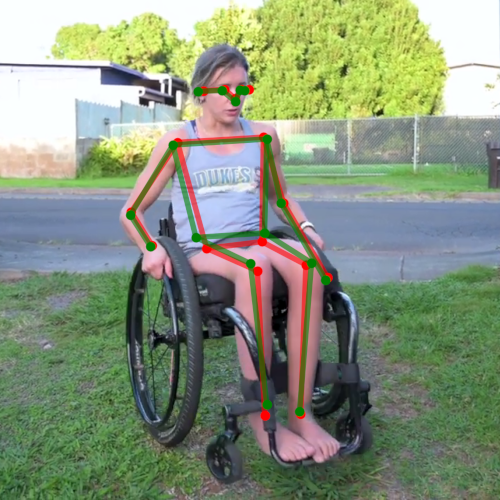}
        \label{subfig:SimilarPerformance4}

        \Description{A wheelchair user is stationary in front of a camera on a field and in front of the road. Both hands are placed on the wheels. The user is not wearing shoes.}
    }
    \vspace{-0.5cm}
    \caption{Examples of similar performance between \systemname{}-Opt and ImageNet in front-facing wheelchair users. Red represents ImageNet predictions while green represents \systemname{}-Opt predictions. \Cref{subfig:SimilarPerformance1,subfig:SimilarPerformance2,subfig:SimilarPerformance3,subfig:SimilarPerformance4} all depict examples of front-facing wheelchair users in a variety of different settings. The \systemname{}-Opt fine-tuned model and ImageNet both performed similarly, generating relatively matching keypoint predictions.}
    \label{fig:SimilarPerformance}
    \vspace{-0.1cm}
\end{figure*}

\noindpar{Similar performance in front-facing scenarios.}
In front-facing scenarios, ImageNet and other pose estimation models often perform very well on wheelchair users. This is because, at this angle, the user can simply be interpreted to be sitting, with all limbs in full view of the camera. Thus in practice, the front-facing wheelchair user is very similar to a front-facing able-bodied user that is sitting. We find that fine-tuning with \systemname{}-Opt performs comparably with the base ImageNet models in front-facing scenarios. Thus, our system maintains crucial information learned from the initial training of ImageNet that has proven to work well on wheelchair users already. Examples of this phenomenon are shown in \Cref{fig:SimilarPerformance}.

\begin{figure*}
    \subfigure[]{
        \includegraphics[width=.23\linewidth]{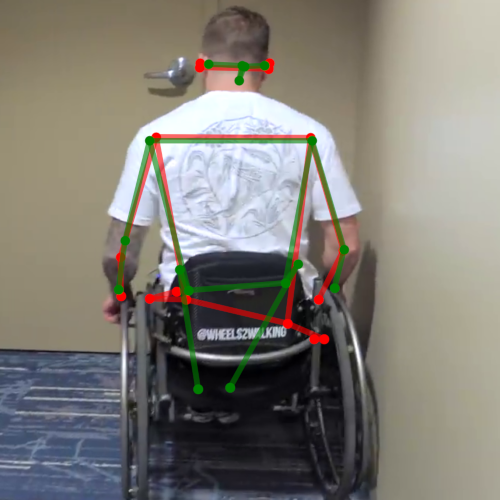}
        \label{subfig:SelfOcclusionImprovements1}

        \Description{A wheelchair user near the corner of the room with their back to the camera. Both hands are placed on the wheels of the wheelchair.}
    }
    \subfigure[]{
        \includegraphics[width=.23\linewidth]{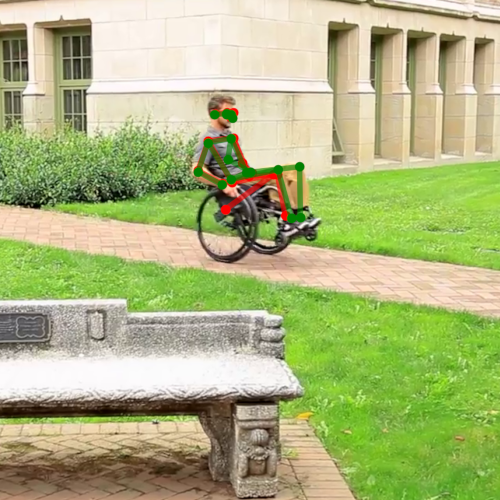}
        \label{subfig:SelfOcclusionImprovements2}

        \Description{A side-view of a wheelchair user traveling down an outdoor path. Both hands are on the wheels doing a slight wheelie.}
    }
    \subfigure[]{
        \includegraphics[width=.23\linewidth]{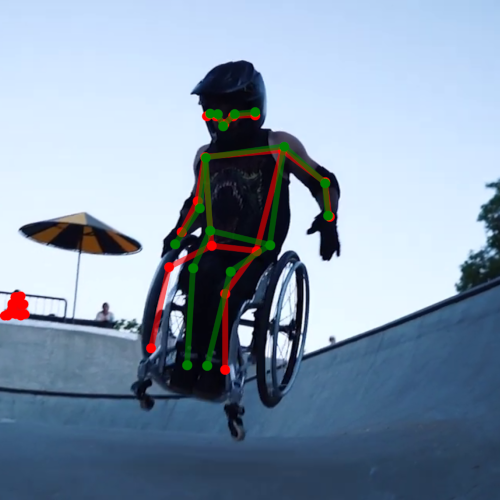}
        \label{subfig:SelfOcclusionImprovements3}

        \Description{A wheelchair athlete at the skate park. The athlete is at a 45-degree angle to the camera in the middle of the air. They are wearing protective gear including a helmet and gloves.}
    }
    \subfigure[]{
        \includegraphics[width=.23\linewidth]{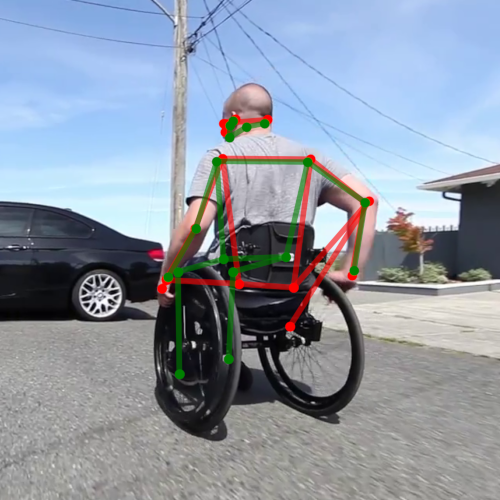}
        \label{subfig:SelfOcclusionImprovements4}

        \Description{
            A wheelchair user on the street traveling at a 45-degree angle away from the camera towards a parked car. Both hands are placed on the wheels as they move towards the car.
        }
    }
    \vspace{-0.5cm}
    \caption{Examples of improvements in keypoint estimation in \systemname{}-Opt over ImageNet in scenarios where the wheelchair occludes part of the user's body. Red represents ImageNet predictions while green represents \systemname{}-Opt predictions. \Cref{subfig:SelfOcclusionImprovements1,subfig:SelfOcclusionImprovements2,subfig:SelfOcclusionImprovements3,subfig:SelfOcclusionImprovements4} all display improvements from \systemname{}-Opt on keypoint predictions, specifically in the lower body.}
    \label{fig:SelfOcclusionImprovements}
     \vspace{-0.3cm}
\end{figure*}

\noindpar{Improvements in keypoint estimation in wheelchair self-occlusion scenarios.}
While existing pose estimation models may work well when the wheelchair user is facing directly forward, they often break down when the user is turned away and the wheelchair begins obstructing the view of the full human body. We find that the additional synthetic data from \systemname{}-Opt helps Detectron2 discern between what is a part of the wheelchair and what is a part of the user's body for a more accurate prediction. \Cref{fig:SelfOcclusionImprovements} illustrates a few examples of such poor performance in ImageNet and improved predictions enabled through \systemname{}-Opt. In situations where the legs are fully occluded by the wheelchair like in \Cref{subfig:SelfOcclusionImprovements1,subfig:SelfOcclusionImprovements4}, our system generates more reasonable predictions compared to those of ImageNet, which placed the legs onto the wheels of the wheelchair or even the elbow. \Cref{subfig:SelfOcclusionImprovements2} shows an example of how \systemname{}-Opt can improve the detection of self occluded keypoints like the user's left knee and ankle. \Cref{subfig:SelfOcclusionImprovements2,subfig:SelfOcclusionImprovements3} shows examples of where ImageNet has mistakenly classified parts of the wheelchair as a keypoint. 

\begin{figure*}
    \subfigure[]{
        \includegraphics[width=.23\linewidth]{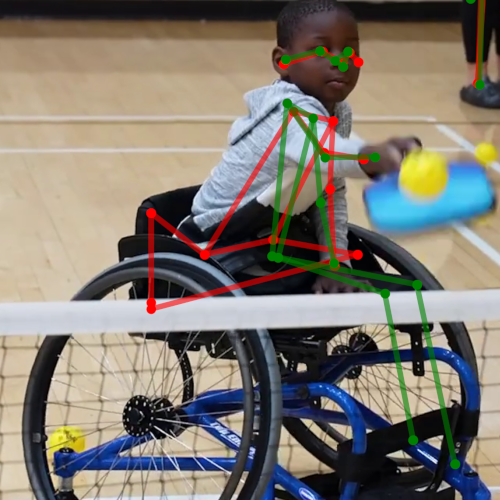}
        \label{subfig:OverfittingWheelchairs1}

        \Description{A young child without lower limbs at a 45-degree angle in a wheelchair playing pickleball. The child is partially occluded by the pickleball net.}
    }
    \subfigure[]{
        \includegraphics[width=.23\linewidth]{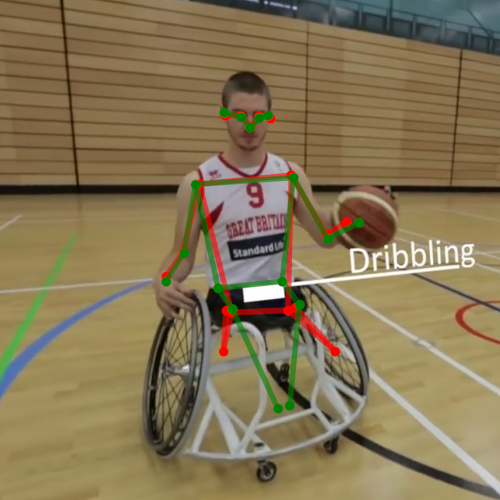}
        \label{subfig:OverfittingWheelchairs2}

        \Description{A wheelchair basketball player without lower limbs is dribbling while facing the camera. The players right hand is on the wheel and the left hand is dribbling the basketball.}
    }
    \subfigure[]{
        \includegraphics[width=.23\linewidth]{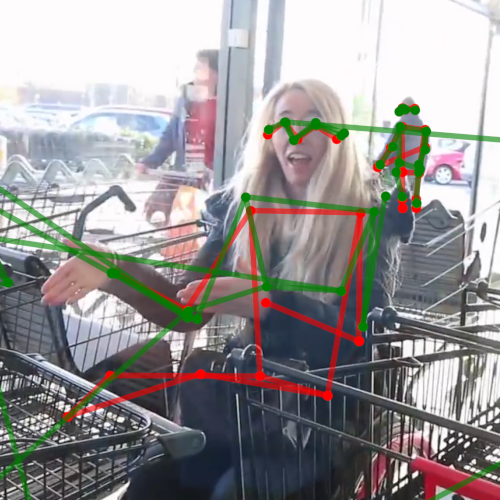}
        \label{subfig:OverfittingWheelchairs3}

        \Description{A person is at a 45-degree angle in a motorized grocery cart behind other grocery carts. They are explaining something with both hands sticking out.}
    }
    \subfigure[]{
        \includegraphics[width=.23\linewidth]{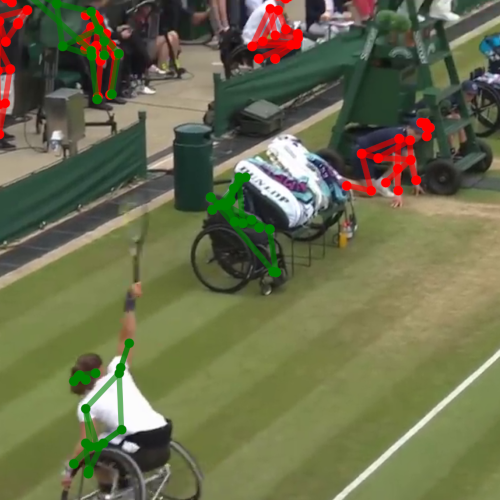}
        \label{subfig:OverfittingWheelchairs4}

        \Description{A game of wheelchair tennis. The camera is at a 45-degree angle in an overhead view. A wheelchair user is raising their racket up in front of an empty wheelchair and a person crouches on the ground. There is a crowd of spectators beyond the field of the game.}
    }
    \vspace{-0.5cm}
    \caption{Examples of overfitting on wheelchairs from \systemname{}-Opt. Red represents ImageNet predictions while green represents \systemname{}-Opt predictions. \Cref{subfig:OverfittingWheelchairs1,subfig:OverfittingWheelchairs2} Examples of poor predictions on wheelchair users without lower limbs. \Cref{subfig:OverfittingWheelchairs3} An example of overfitting onto any wheelchair resembling object, like a grocery cart. \Cref{subfig:OverfittingWheelchairs4} An example of overfitting where even an empty wheelchair is detected to have a user.}
    \label{fig:OverfittingWheelchairs}
     \vspace{-0.2cm}
\end{figure*}

\noindpar{Overfitting on wheelchairs.}
Upon examining the predictions made by ImageNet and ImageNet fine-tuned with \systemname{}-Opt, we notice that both systems perform poorly on users with no lower limbs. As shown in \Cref{subfig:OverfittingWheelchairs1,subfig:OverfittingWheelchairs2} While ImageNet tends to classify the wheelchair as the missing legs, we notice that our system instead "fills in the blanks" and predicts legs in reference to the wheelchair where they might usually be for a wheelchair user. We further notice that our system tends to have more false positives in detecting what can be defined as a wheelchair. As seen in \Cref{subfig:OverfittingWheelchairs3}, objects that resemble a wheelchair, like a grocery cart, may affect the keypoint estimation of a wheelchair user. In other cases like the one shown in \Cref{subfig:OverfittingWheelchairs4}, we see that our system can even detect empty wheelchairs and fill them in with humans when there are not any. While this shows great promise in the information transfer between digitally modeled mobility assistive devices and real-world data, we find these tendencies can obfuscate the real postures of wheelchair users.

\section{Discussion and Future Work}
\label{sec:Discussion}

\changed{\noindpar{Generalizable knowledge and transferability to HCI research.} Data generation approaches have been widely adopted in projects in HCI for problems involving thermal imaging \cite{hu2020fingertrak}, IMU \cite{Jain2023VirtualIMUEating, Rey2019IMUTraining}, stroke gesture \cite{Leiva2017SynthesizingStrokeGestures}, and RF data \cite{Ahuja2021Vid2Doppler, Cai2020RFSynth}. Given the popularity of data generation in HCI, we believe our techniques could be easily transferable to related and future works in accessibility, motion generation, and pose estimation. Furthermore, the modularization in our pipeline could improve transferability by facilitating segmented changes -- a flexible way for data synthesis to experiment with different components. Finally, methods shown in our statistical analysis and model performance evaluation could be highly reusable in future work that adopts our technique. That being said, our pipeline provides the baseline framework for futures efforts in  research that require different humanoid models, motion synthesis techniques for upper body and lower body, environmental factors, and VR toolchains.} 

\changed{\noindpar{Configurability to cater needs of developers and end users.} We believe the results shown in \Cref{tab:Ablation}, \Cref{sec:ClipFix}, and \Cref{sec:KeypointLocation} demonstrate a clear need for developers to have an interactive tool to modify training data, where slight changes in the data modeling can have major effects on AI performance. Currently, there are few tools to do so, with many developers choosing to leverage static motion datasets which may not perfectly fit their needs. This has been one of the dominant reasons for the rise of inequitable AI models. For this reason, \systemname{} is a pipeline specifically designed to enable a high degree of configurability. This allows for the creation of personalized synthetic datasets that cater to developers' needs, thereby increasing the likelihood of catering to the needs of end users. This would lead to more effective and inclusive AI models, which can be tuned to the individual needs of the user instead of a one-size-fits-all solution for better performance in real-world applications.}


\changed{\noindpar{Improvements on the realism of generated data.} We hope the use of Unity enables future research into this idea and enables developers to build a synthetic data generator that extends beyond our simple simulation environment using 3D modeled scenes and rooms instead of flat background images. Recent achievements in Neural Radiance Fields (NeRF) could be leveraged to synthesize photorealistic background images that adapt in response to changes in the virtual camera's perspective. Future work in realistic 3D environment modeling will enable research in surface semantics for more realistic configurations -- having wheelchair users positioned at ground surfaces. Moreover, physics could be incorporated to simulate the locomotion of wheelchair users which would enable the modeling of more realistic motions and collisions with other models compared to our simple animations.}



\changed{\noindpar{Efficacy of human evaluation.} We found that motions from motion generation models and motion capture were largely perceived similarly by participants in the human evaluation (\Cref{sec:HumanEvaluation}). We further found that these two types of motions after being filtered for human-perceived realism performed comparably in model performance evaluation (\Cref{tab:Ablation}). 
However, whether due to current generative AI performance or the lack of training data for disability-related movement, our human evaluations indicate that generative AI models are still not able to represent users with disabilities accurately without external human feedback.
This is shown in the improved performance with the addition of human evaluations (\Cref{tab:Ablation}). 
Due to this, we believe that human evaluators are still vital to ensure that generated data is representative. We recognize that our method for assessing motions, although not exhaustive or entirely free of bias, contributes as an initial stride towards creating datasets that are more inclusive.
Future research should extend upon this work to integrate a more involved human-centered system which will, instead, allow evaluators to meaningfully guide the generation of data, including motions, simulation scenarios, and user modeling, through iterative feedback (e.g., guided prompting) to create a more representative dataset. We also found little literature on motion synthesis for people with disabilities and recognized this vacuum as both a challenge and an opportunity for generative AI.}

\changed{\noindpar{Comprehensive model performance evaluation.}} Our model testing did not involve a serious grid search for data generation hyperparameters, model seeds, initializations, or model configurations. We also used the same training strategy for all tested datasets. We held all these values constant to focus on the quality and impact of the synthetic dataset on model performance. Even with this naïve approach to training, \systemname{} still yielded promising results in the improvement of pose estimation models through synthetic data on wheelchair users. We found noticeable improvements in both person detection and pose estimation problems. This indicates that the specific synthetic modeling of users in wheelchairs in \systemname{} can make existing computer vision models more equitable by improving performance on wheelchairs. We believe our findings pave the way for future works in synthetic data on humans with other mobility-assistive technologies to improve pose estimation equitability.

\changed{\noindpar{Diverse participant groups.} A key limitation in this work is that we only analyze the digitalized representation of users with all four limbs and feet fixed on the foot rest, which does not express the full range of wheelchair users with different bodies such as those with amputations, dwarfism, spinal deformities, and other conditions. Thus the findings may not be reflective of the wheelchair population at large. We hope our data generation pipeline provides a framework and will enable future developers to leverage 3D modeling and Unity tools to create a more diverse body of wheelchair user models. For instance, developers can easily add new bones to existing models to more realistically represent spinal deformities. We hope that these tools will also enable future works analyzing different disabilities and mobility assistive devices which were not addressed in our current research.}

\changed{\noindpar{Improve inclusiveness of AI for more recognition modalities.}} Furthermore, \systemname{} enables more work beyond 2D bounding boxes and pose estimation. Annotations on depth, surface normals, object segmentation, occlusion, 3D bounding boxes, and 3D keypoints are fully implemented in our current data synthesis pipeline but still unexplored. These annotations can be even more difficult and costly to collect compared to RGB images and 2D pose annotations, often requiring the use of specialized equipment and data collection processes. Synthetic data has no such problem, where any desired annotation and labeling are all equally accessible to collect. Thus, we believe that \systemname{} can be adapted to potentially address problems in wheelchair pedestrian detection with object segmentation and occlusion annotations \cite{dollar2011pedestrian}, 3D pose estimation using 3D bounding box and pose annotations, and robotics detection of people and mobility aids through depth data \cite{vasquez2017deep} among other accessibility-related problems cheaply and efficiently.

\changed{\noindpar{Pitfalls of exclusion in data generation vs. data collection.} \systemname{} is a pipeline for both AI developers and wheelchair users to circumvent existing inaccessible data collection methods and meaningfully improve the training process of AI models by generating data for wheelchair users. 
However, we are cautious that by circumventing existing inaccessible data collection methods with our tool, we could run the potential risk of furthering exclusion, which echoes long-standing debates within the accessibility community. Our paper is based on the assumption that making AI equitable requires pursuing multiple approaches together -- effective approaches to improving representations of training data from people with disabilities leveraging both data collection and data generation. 
We advocate that new tools for data generation and the existing data collection methods are not mutually exclusive. Their synergy could lead to a more practical approach to resolving accessibility challenges than what could be offered by either of the two approaches alone. We believe that the following characteristics are vital in future works to avoid pitfalls of exclusion: 1) representative and diverse participant groups, for which we have conducted studies around spinal injuries of various levels and recommend future work to consider participants from wider backgrounds; 2) realistic generated data, for which we invented several data generation techniques optimizing data realism; and 3) effective tools for human evaluation, for which we adopted embodiment in our human evaluation interface allowing participants to seamlessly transfer the presented motion sequences to their own bodies for a more intuitive evaluation.}

\section{Conclusion}

We introduce \systemname{}, an extension of the highly-parameterized synthetic data generator PeopleSansPeople, for wheelchair users with the possibility of use in other mobility-assistive technologies to improve the performance of common pose estimation algorithms in the traditionally underrepresented group of wheelchair users. \systemname{} includes a full end-to-end pipeline to convert existing motion capture and motion generation model outputs into wheelchair user animations for use in a complete Unity Simulation scene \textbf{(RQ1.1)} containing a range of 3D human models from Unity SyntheticHumans in wheelchairs, backgrounds, occluders, and unique lighting conditions. We provide full control over all related parameters, including keypoint labeling schema, for computer vision tasks \textbf{(RQ1.2)}. We tested our pipeline using two different motion sequence sources: motion capture data from HumanML3D and motion generation outputs from Text2Motion. These motions underwent a set of human evaluations. We then analyzed the impacts of different domain randomization parameters and motions on model performance, finding an "optimal" combination of parameters and comparable performance between our motion sources \textbf{(RQ2.1)}. Finally, we tested the model performance of the dataset generated through \systemname{} with no real-world data using the optimal parameters found previously on a dataset of real wheelchair users to find noticeable improvements in model performance when compared against existing pose estimation models \textbf{(RQ2.2)}. We expect \systemname{} to enable a new range of research in using synthetic data to model users with disabilities in improving the equity of AI.

\balance
\bibliographystyle{ACM-Reference-Format}
\bibliography{ref.bib}

\appendix
\newpage
\section{Appendix}

\renewcommand\figurename{Appendix figure}
\crefalias{figure}{appendixfigure}
\setcounter{figure}{0}

\renewcommand\tablename{Appendix table}
\crefalias{table}{appendixtable}
\setcounter{table}{0}

\subsection{Animation Clip Fix}
\label{sec:ClipFix}
\begin{figure*}[ht]
    \includegraphics[width=\linewidth]{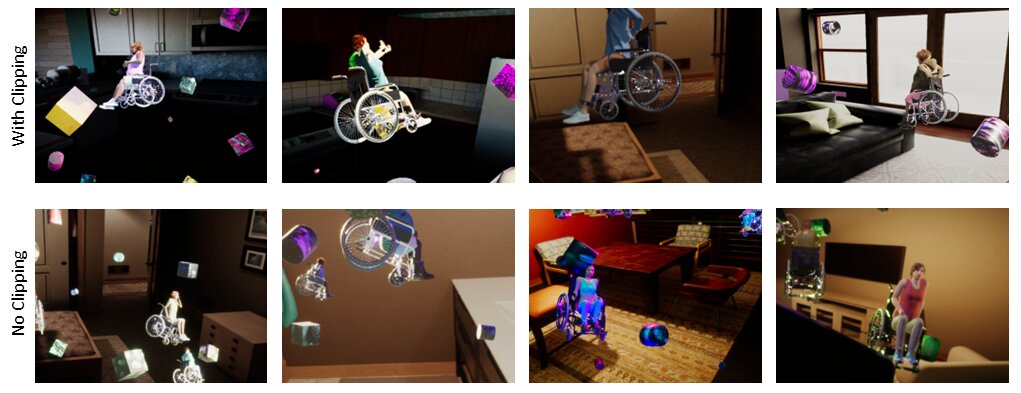}
    \caption{Examples of generated data with and without increased lower body clipping. Notice the overlap between the ankles and the footplate in the motion sequences with clipping creates models where it looks like the human model has fused into the wheelchair.}
    \label{fig:ClippingExamples}

    \Description{Four examples of animations with clipping and four examples of animations without clipping. Examples of clipping feature the human model with feet that extend into the footplate of the wheelchair and below. Examples without clipping place the feet on top of the footplate.}
\end{figure*}

To ensure the human model's do not overlap with the wheelchair footplates, a $35$ degree rotation of the hips up toward the sky is applied. This hip rotation decreases the amount of clipping between the legs and the wheelchair. Clipping is a common occurrence in 3D modeling where when objects are within each other, only the object closer to the camera will be rendered and obscure the overlapped object. Examples of this extra clipping are shown in \cref{fig:ClippingExamples}. 

\subsection{Posture to \textit{AnimationClip} Conversion}

We convert the postures resulting from previous sections into human images using human models in Unity which take \textit{AnimationClips} as input for pose configurations. To convert pose frames into \textit{AnimationClips}, all motion sequences from each set, represented by a series of joint rotations, are individually imported into Blender. Each frame's joint rotations are applied to the corresponding joint in a Unity Perception human model Blender template and exported as an FBX file. Upon importing an FBX file into Unity, Unity will automatically convert all baked animations into Unity-readable \textit{AnimationClip} files which can be used in Unity Perception. All \textit{AnimationClips} are then set to read as Unity Humanoid animations for use in data synthesis.

\subsection{\systemname{} Randomizers}
Unity Perception enables the use of the "randomizer" paradigm to enable users to configure the \textit{domain randomization} of individual parameters 
\cite{borkman2021unity}. \systemname{} uses multiple Unity Perception default randomizers, PSP custom randomizers \cite{ebadiPeopleSansPeopleSyntheticData2022}, and a collection of custom \systemname{} randomizers. It is important to note that many of PSP's custom randomizers have made it into the release version of Unity Perception (1.0.0). We have chosen to maintain the original randomizers used for a direct comparison between data synthesized between PSP and \systemname{}. Like in PSP, our randomizers are regarded as further data augmentation techniques which limits the need for data augmentations during training itself. All randomizers sampled values from a uniform distribution. \Cref{tab:EnvironmentParameters} outlines the statistical distributions for our randomizer parameters. A brief description of each randomizer used in \systemname{} is described below. 

\noindent\textbf{BackgroundObjectPlacementRandomizer.} Randomly spawns background and occluder objects within a user-defined 3D volume. Separation distance can be set to dictate the proximity of objects from each other. Poisson-Disk sampling \cite{bridson2007Poisson} is used to randomly place objects sourced from a set of primitive 3D game objects (cubes, cylinders, spheres, etc.) from Unity Perception in a given area.

\noindent\textbf{BackgroundOccluderScaleRandomizer.} Randomizes the scale of the background and occluder objects. 

\noindent\textbf{RotationRandomizer.} Randomizes the 3D rotation of background and occluder objects.

\noindent\textbf{ForegroundObjectPlacementRandomizer.} Similar to \textit{BackgroundObjectPlacementRandomizer}. Randomly spawns foreground objects selected from the default set of PSP models affixed in wheelchair models. 

\noindent\textbf{ForegroundScaleRandomizer.} Similar to \textit{BackgroundOccluderScaleRandomizer.} Randomizes the scale of foreground objects.

\noindent\textbf{TextureRandomizer.} Randomizes the texture of predefined objects provided as a JPEG or PNG. We used the set of example textures from Unity Perception which are applied to the background and occluder objects as well as to the background wall when no specific background is set.

\noindent\textbf{HueOffsetRandomizer.} Randomizes the hue offset applied to textures on the object. Applied to background and occluder objects as well as to the background wall when no specific background is set. 

\noindent\textbf{SpriteRandomizer.} Randomizes the background wall. Used as an alternative to the \textit{TextureRandomizer} when images should not be stretched to fill a canvas.

\noindent\textbf{HumanGenerationRandomizer.} Randomizes the age, sex, ethnicity, height, weight, and clothing of spawned human assets. Humans are spawned in batches called pools which are periodically regenerated through the simulation process. All humans are spawned within a predefined base which contains the wheelchair model used. All textures and models used are sourced directly from SyntheticHumans.

\noindent\textbf{NonTagAnimationRandomizer.} Randomizes the pose applied to a character. The pose is a randomly selected frame from a randomly selected \textit{AnimationClip} taken from a universal pool of \textit{AnimationClips}. Provides a custom alternative to the Unity Perception AnimationRandomizer for randomizing animations taken from a single pool.

\noindent\textbf{TransformPlacementRandomizer.} Randomizes the position, rotation, and size of generated SyntheticHumans. Rotations around the \textit{X,Z}-axis are limited to better represent real world data where users are rarely seen in such orientations.

\noindent\textbf{SunAngleRandomizer.} Randomizes a directional light's intensity, elevation, and orientation to mimic the lighting effects of the Sun.

\noindent\textbf{LightRandomizer.} Randomizes a light's intensity and color (RGBA). Also enables the randomization of a light's on/off state.

\noindent\textbf{LightPositionRotationRandomizer.} Randomizes a light's global position and rotation in the scene.

\noindent\textbf{CameraRandomizer.} Randomizes the extrinsic parameters of a camera including its global position and rotation. Enables the randomization of intrinsic camera parameters including field of view and focal length to better mimic a physical camera. Adds camera bloom and lens blur around objects that are out of focus to capture more diverse perspectives of the scene.

\noindent\textbf{PostProcessVolumeRandomizer.} Randomizes select post processing effects including vignette, exposure, white balance, depth of field, and color adjustments. 

\begin{table*}[]
\caption{Domain randomization parameters of \systemname{}}
\label{tab:EnvironmentParameters}
\tiny
\renewcommand{\arraystretch}{1.5}
\resizebox{\linewidth}{!}{
\begin{tabular}{l|l|l|l}
\hline
\textbf{Category} &
  \textbf{Randomizer} &
  \textbf{Parameter} &
  \textbf{Distribution} \\ \hline
\multirow{7}{*}{Background/Occluder Objects} &
  \multirow{2}{*}{BackgroundObjectPlacementRandomizer} &
  object placement &
  \texttt{Cartesian{[}Uniform(-7.5, 7.5), Uniform(-7.5, 7.5), Uniform(-7.5, 7.5){]}} \\ \cline{3-4} 
 &
   &
  separation distance &
  \texttt{Cartesian{[}Constant(2.5), Constant(2.5), Constant(2.5){]}} \\ \cline{2-4} 
 &
  BackgroundOccluderScaleRandomizer &
  object scale range &
  \texttt{Cartesian{[}Uniform(1, 12), Uniform(1, 12), Uniform(1, 12){]}} \\ \cline{2-4} 
 &
  RotationRandomizer &
  object rotation &
  \texttt{Euler{[}Uniform(0, 360), Uniform(0, 360), Uniform(0, 360){]}} \\ \cline{2-4} 
 &
  TextureRandomizer &
  textures &
  A set of of texture assets \\ \cline{2-4} 
 &
  SpriteRandomizer &
  sprites &
  A set of sprite assets \\ \cline{2-4} 
 &
  HueOffsetRandomizer &
  hue offset &
  \texttt{Uniform(-180, 180)} \\ \hline
\multirow{16}{*}{Human Model} &
  \multirow{8}{*}{HumanGenerationRandomizer} &
  humans per iteration &
  \texttt{Uniform(5, 12)} \\ \cline{3-4} 
 &
   &
  human pool size &
  \texttt{Constant(50)} \\ \cline{3-4} 
 &
   &
  pool refresh interval &
  \texttt{Constant(400)} \\ \cline{3-4} 
 &
   &
  age &
  \texttt{Uniform(10, 100)} \\ \cline{3-4} 
 &
   &
  height &
  \texttt{Uniform(0.1, 1)} \\ \cline{3-4} 
 &
   &
  weight &
  \texttt{Uniform(0, 1)} \\ \cline{3-4} 
 &
   &
  sex &
  male, female \\ \cline{3-4} 
 &
   &
  ethnicity &
  Caucasian, Asian, Latin American, African, Middle Eastern \\ \cline{2-4} 
 &
  \multirow{3}{*}{TransformPlacementRandomizer} &
  synthetic human placement &
  \texttt{Cartesian{[}Uniform(-7.5, 7.5), Uniform(-7.5, 7.5), Uniform(-4, 1){]}} \\ \cline{3-4} 
 &
   &
  synthetic human rotation &
  \texttt{Euler{[}Uniform(0, 20), Uniform(0, 360), Uniform(0, 20){]}} \\ \cline{3-4} 
 &
   &
  synthetic human size range &
  \texttt{Cartesian{[}Uniform(0.5, 3), Uniform(0.5, 3), Uniform(0.5, 3){]}} \\ \cline{2-4} 
 &
  \multirow{2}{*}{ForegroundObjectPlacementRandomizer} &
  predefined model placement &
  \texttt{Cartesian{[}Uniform(-7.5, 7.5), Uniform(-7.5, 7.5), Uniform(-9, 6){]}} \\ \cline{3-4} 
 &
   &
  predefined model separation distance &
  \texttt{Cartesian{[}Constant(3), Constant(3), Constant(3){]}} \\ \cline{2-4} 
 &
  ForegroundScaleRandomizer &
  predefined model scale range &
  \texttt{Cartesian{[}Uniform(0.5, 3), Uniform(0.5, 3), Uniform(0.5, 3){]}} \\ \cline{2-4} 
 &
  ForegroundRotationRandomizer &
  predefined model rotation &
  \texttt{Euler{[}Uniform(0, 20), Uniform(0, 360), Uniform(0, 20){]}} \\ \cline{2-4} 
 &
  NonTagAnimationRandomizer &
  animations &
  A set of AnimationClips of arbitrary length \\ \hline
\multirow{8}{*}{Lights} &
  \multirow{3}{*}{SunAngleRandomizer} &
  hour &
  \texttt{Uniform(0, 24)} \\ \cline{3-4} 
 &
   &
  day of the year &
  \texttt{Uniform(0, 365)} \\ \cline{3-4} 
 &
   &
  latitude &
  \texttt{Uniform(-90, 90)} \\ \cline{2-4} 
 &
  \multirow{3}{*}{LightRandomizer} &
  intensity &
  \texttt{Uniform(5000, 50000)} \\ \cline{3-4} 
 &
   &
  color &
  \texttt{RGBA{[}Uniform(0, 1), Uniform(0, 1), Uniform(0, 1), Constant( 1){]}} \\ \cline{3-4} 
 &
   &
  enabled &
  $P(enabled)=0.8, P(disabled)=0.2$ \\ \cline{2-4} 
 &
  \multirow{2}{*}{LightPositionRotationRandomizer} &
  position offset from initial position &
  \texttt{Cartesian{[}Uniform(-3.65, 3.65), Uniform(-3. 65, 3.65), Uniform(-3.65,   3.65){]}} \\ \cline{3-4} 
 &
   &
  rotation offset from initial rotation &
  \texttt{Euler{[}Uniform(-50, 50), Uniform(-50, 50), Uniform(-50, 50){]}} \\ \hline
\multirow{4}{*}{Camera} &
  \multirow{4}{*}{CameraRandomizer} &
  field of view &
  \texttt{Uniform(5, 50)} \\ \cline{3-4} 
 &
   &
  focal length &
  \texttt{Uniform(1, 23)} \\ \cline{3-4} 
 &
   &
  position offset from initial position &
  \texttt{Cartesian{[}Uniform(-5, 5), Uniform(-5, 5), Uniform(-5, 5){]}} \\ \cline{3-4} 
 &
   &
  rotation offset from initial rotation &
  \texttt{Euler(Uniform(-5, 5), Uniform(-5, 5), Uniform(-5, 5){]}} \\ \hline
\multirow{6}{*}{Post Processing} &
  \multirow{6}{*}{PostProcessVolumeRandomizer} &
  vignette intensity &
  \texttt{Uniform(5, 50)} \\ \cline{3-4} 
 &
   &
  fixed exposure &
  \texttt{Uniform(5, 10)} \\ \cline{3-4} 
 &
   &
  white balance temperature &
  \texttt{Uniform(-20, 20)} \\ \cline{3-4} 
 &
   &
  depth of field focus distance &
  \texttt{Uniform(.1, 4)} \\ \cline{3-4} 
 &
   &
  color adjustments: contrast &
  \texttt{Uniform(-30, 30)} \\ \cline{3-4} 
 &
   &
  color adjustments: saturation &
  \texttt{Uniform(-30, 30)} \\ \hline
\end{tabular}}
\end{table*}

\subsection{Testing Dataset Action Classes}
A set of real-world wheelchair data was collected from \textit{YouTube}. A predefined set of $16$ action classes was defined before being used as keyword searches to identify relevant videos. Action classes were selected based on a mix of common actions and unique wheelchair movements. A total of $2,464$ images were collected. More information on action classes is found in \cref{tab:ActivityClasses}.

\begin{table*}[ht]
\caption{Distribution of activity classes in the testing dataset.}
\label{tab:ActivityClasses}
\begin{tabular}{@{}ll@{}}
\toprule
\textbf{Activity Class}    & \textbf{Percentage of Dataset} \\ \midrule
talking           & 21.659\%              \\
wheelchair skills & 14.692\%              \\
daily routine     & 13.460\%              \\
dance             & 10.664\%               \\
basketball        & 8.863\%              \\
tennis            & 5.829\%               \\
extreme sports    & 5.640\%               \\
general sports    & 4.645\%               \\
household chores  & 3.318\%               \\
shopping          & 2.180\%               \\
cooking           & 2.085\%               \\
travel            & 1.801\%                \\
photoshoot        & 1.611\%               \\
rugby             & 1.422\%               \\
pickleball        & 1.185\%               \\
stretches         & 0.948\%               \\ \bottomrule
\end{tabular}
\end{table*}

\subsection{Impacts of Keypoint Location Definitions}
\label{sec:KeypointLocation}
\begin{figure*}[ht]
    \subfigure[]{
        \includegraphics[width=.23\linewidth]{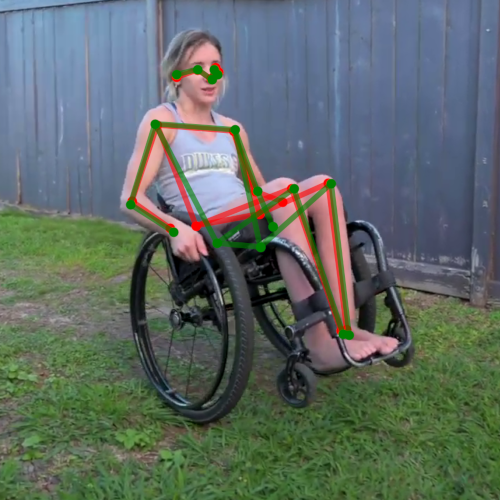}
        \label{subfig:HipAnnotation1}

        \Description{A person in a wheelchair doing a wheelie on grass facing the camera at a 45 degree angle. Both hands are on the wheels and both feet are in the footplate.}
    }
    \subfigure[]{
        \includegraphics[width=.23\linewidth]{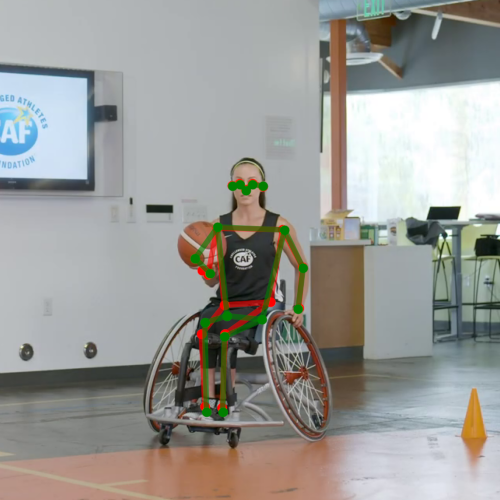}
        \label{subfig:HipAnnotation2}

        \Description{A person in a wheelchair holding a basketball facing the camera. the right hand is holding the basketball near the chest while the left hand is on the wheel.}
    }
    \subfigure[]{
        \includegraphics[width=.23\linewidth]{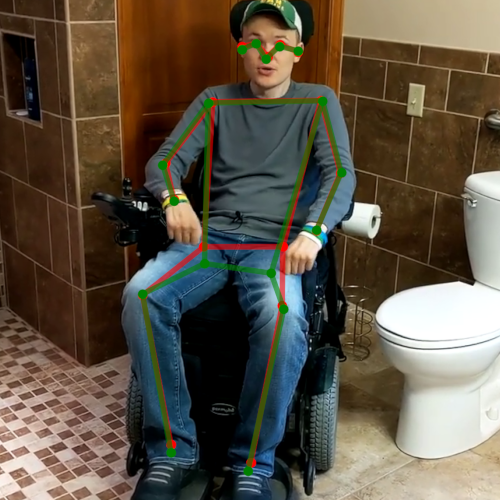}
        \label{subfig:HipAnnotation3}

        \Description{A person in a motorized wheelchair facing the camera in a bathroom. Both elbows rest on the armrest while the arms rest near their lap.}
    }
    \subfigure[]{
        \includegraphics[width=.23\linewidth]{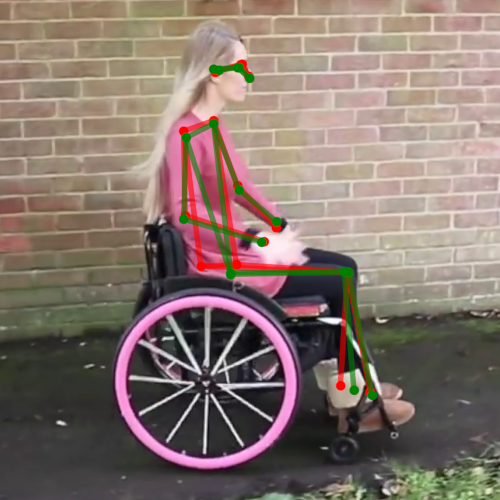}
        \label{subfig:HipAnnotation4}

        \Description{A side view of a person in a wheelchair rolling forward. Both arms are near the person's lap.}
    }

    \caption{Examples of the different prediction outputs between a lower hip definition and a higher hip definition in synthetic data. Green represents the lower hip definition and red represents the higher hip definition. \cref{subfig:HipAnnotation1,subfig:HipAnnotation2,subfig:HipAnnotation3,subfig:HipAnnotation4} all show examples of where other keypoint predictions are relatively similar with the exception of the hips where the lower hip annotations are placed lower on the body compared to the higher hip annotation. Each image depicts the wheelchair user in a different angle, setting, and action.}
    \label{fig:HipAnnotationChanges}
\end{figure*}

Unity Perception enables users to redefine different keypoint definitions for image annotations. Unity Perception provides a default COCO $17$-keypoint annotation schema which places each keypoint directly on the joint between two bones. However, in human-annotated datasets, many evaluators place the hip much higher than the actual joint between the hips and the femur. We test Unity's default keypoint schema with lower hips against our own custom COCO $17$-keypoint schema which raises the hip keypoint up the torso by $\approx9$cm. All data is generated in the exact same way with the only difference of how the hip keypoints are defined. Examples of how the annotation schema affects predictions are displayed in \cref{fig:HipAnnotationChanges}. We see that changes in the definition of keypoints in synthetic data can drastically change the position of the changed keypoint in predictions. We believe this concept can be used to adapt and tune existing pose estimation models with different keypoint definitions through the use of only synthetic data.

\end{document}